\documentclass[letterpaper,twocolumn,10pt]{article}
\usepackage{usenix2019_v3}

\usepackage{filecontents}
\usepackage[]{hyperref} 
\usepackage{epsfig,endnotes}
\usepackage[english]{babel}
\usepackage{blindtext}
\usepackage[normalem]{ulem}
\usepackage{graphicx}
\usepackage{subcaption}
\usepackage{pifont}
\usepackage{amsmath}
\usepackage{xspace}
\usepackage{amssymb} 
\usepackage{fancyhdr}
\usepackage{lipsum}
\usepackage{hyperref}
\usepackage{tikz}
\usepackage{algorithm}
\usepackage{algpseudocode}
\usepackage{booktabs}
\usepackage{needspace}
\usepackage{float}
\usepackage{placeins}
\usepackage{multirow}
\usepackage{xcolor}
\usepackage{enumitem}
\usepackage{setspace}

\usepackage[small, compact]{titlesec}
\usepackage[format=plain,labelfont=bf,font=small]{caption}

\algrenewcommand\algorithmicrequire{\textbf{Input:}}
\algrenewcommand\algorithmicensure{\textbf{Output:}}

\definecolor{mygreen}{HTML}{00c418}
\definecolor{cherryblossom}{HTML}{F891BB}
\definecolor{khaki}{HTML}{41431B}
\definecolor{olivegreen}{HTML}{313E17}
\definecolor{sand}{HTML}{BFA28C}
\definecolor{coralblue}{HTML}{3852B4}
\definecolor{oceanblue}{HTML}{5478FF}
\definecolor{leafgreen}{HTML}{519A66}
\definecolor{darkgray}{HTML}{605B51}

\newcommand{\sysname}{Lodestar\xspace}
\newcommand{\sysnameshort}{LS\xspace}
\newcommand{\gatewayfull}{Stateful Gateway\xspace}
\newcommand{\ras}{Routing Service\xspace}
\newcommand{\aibrix}{AIBrix\xspace}
\newcommand{\sglang}{SGLang\xspace}

\newcommand{\heuristic}{Prefix-cache-and-load-aware\xspace}
\newcommand{\prefixaware}{Prefix-cache\xspace}

\makeatletter
\renewcommand\@maketitle{\newpage
  \vbox{%
    \vskip 2em
    \begin{center}%
      {\Large\bf \@title \par}%
      \vskip 0.2in minus 0.15in
      {\large\it
        \lineskip .5em
        \begin{tabular}[t]{c}\@author
        \end{tabular}\par}%
    \end{center}%
    \par
    \vskip 1.5em
  }%
}
\makeatother

\begin{document}

\pagestyle{empty} 
\thispagestyle{empty}



\date{}

\title{\Large \bf \sysname: An Online-Learning LLM Inference Router\xspace}

\author{
\begin{tabular}{ccc}
{\rm Gangmuk Lim$^{\dagger}$} & {\rm Wanyu Zhao} & {\rm Brighten Godfrey} \\
UIUC & UIUC & UIUC \\
gangmuk2@illinois.edu & wanyu2@illinois.edu & pbg@illinois.edu \\[1ex]
{\rm Jiaxin Shan} & {\rm Le Xu} & {\rm Liguang Xie} \\
Bytedance & University of Edinburgh & Bytedance \\
seedjeffwan@gmail.com & le.xu@ed.ac.uk & liguang.xie@bytedance.com
\end{tabular}
}

\maketitle

\renewcommand{\thefootnote}{$\dagger$}
\footnotetext{Corresponding author.}
\renewcommand{\thefootnote}{\arabic{footnote}}


\begin{abstract}


Efficiently serving large language model (LLM) inference tasks is crucial both for user-perceived latency such as time-to-first-token (TTFT) and for GPU utilization. However, LLM request routing---that is, assigning each inference request to a GPU instance -- is particularly challenging: execution is highly input-dependent; batching and KV-cache reuse create strong cross-request coupling; and latency responds nonlinearly to context length, model/engine settings, and heterogeneous accelerators. As a result, simple traditional load balancing algorithms, and even heuristics tailored for LLM inference, fail to achieve good performance. We present \sysname, a novel learning-based request routing system for distributed GPU clusters. \sysname continuously collects a snapshot of the cluster at per-request level---including real-time instance state, request characteristics, and observed performance---and trains an online reward predictor that it uses to route inference requests to the instance that will maximize given reward (e.g., minimizing TTFT). \sysname is cloud-native and works seamlessly with existing serving stacks (vLLM). With continuous online adaptation to changing workloads and infrastructure conditions, \sysname achieves 1.41$\times$ lower average TTFT and 1.47$\times$ lower P99 TTFT on average (up to 2.15$\times$/1.86$\times$ on homogeneous and 4.38$\times$/4.42$\times$ on heterogeneous clusters) compared to a state-of-the-art prefix cache and load-aware heuristic, and learns these efficient routing strategies within $\approx$5 minutes, based on experiments in a public cloud GPU cluster.



\end{abstract}

\section{Introduction}
\label{sec:intro}

The deployment of large language models at scale has created unprecedented challenges in optimizing distributed systems to support them. LLM inference workloads require fast response to queries as they are part of interactive applications~\cite{ chatbotarena24chiang}, but they have drastically different characteristics than traditional latency-sensitive applications such as CPU-based user-facing cloud services and various supporting services like key-value stores and databases. LLM inference consumes much more compute, runs at a higher cost, has much higher latency, and has higher variance in resource consumption and latency between different requests. 
In traditional web applications, one request may take tens of milliseconds of CPU; a single LLM inference may easily run for a few seconds to tens of seconds, on an expensive GPU~\cite{deepspeedinference, fastertransformer, turbotransformers}. Even within the LLM inference workloads, different workloads exhibit highly variable computational requirements, ranging from short conversational exchanges~\cite{zheng2024lmsyschatm} to long document understanding~\cite{hsieh2024ruler} to complex reasoning tasks~\cite{jaech2024openai, guo2025deepseek} involving highly dynamic input and output token ranges.

Our work concerns how to map each incoming inference request to one of the many instances (or ``pods'' in Kubernetes) within a cluster, which we refer to as \emph{inference routing}. Inference routing has a significant impact on LLM inference performance. Traditional load balancing techniques, like round robin or variants of Join the Shortest Queue (JSQ)~\cite{ptwo01mitzenmacher}, seek to balance queue length but could result in significant inefficiency and under-utilization, because they do not take into account that there will be significant differences in the resources needed to run different requests and also to run an inference request on different instances at different moments in time. In particular, some instances may have cached the state from processing prefixes of an inference query, allowing it to reuse large portions of work~\cite{vllm, sglang, marconi, promptcache}.  Recent LLM serving infrastructures have introduced more sophisticated LLM-specific routing strategies that consider such prefix sharing and cache locality~\cite{aibrix, dynamo, llm-d, preble}. 
While this offers an improvement for LLM inference routing compared to generic load balancing, it is hard for manually written heuristics or relatively simple cost models to capture all the important nuances that can affect inference performance---such as multi-dimensional resources (GPU, memory, CPU), dramatic heterogeneity in resources across workloads and requests within a workload, differences in resources across multiple GPU models within a cluster, and tradeoffs between prefix cache opportunity vs. load to optimize the latency. Simultaneously, the underlying serving infrastructure experiences rapid state changes due to GPU compute/memory utilization patterns, the number of running tokens for prefill and decode phase, key-value (KV) cache hit ratios, dynamic request concurrency, and queue length.




We argue that the combination of these subtleties and the high expense of serving LLM workloads means a new approach is needed. This paper introduces \sysname, an online learning-based, cloud-native distributed request routing system for LLM inference. 
\sysname collects detailed cluster states and aims to leverage them with the power of deep learning to navigate the complex environment. At a high level, \sysname learns to accurately predict latency 
that would result from routing a particular inference request to a particular instance, given the request information and cluster state at the time of routing.
Then, it applies this predictor to route requests to the instance that maximizes predicted reward such as negative of time-to-first-token (TTFT). 
While this idea is straightforward, we need to solve two key challenges. 

\paragraph{1. Accurate online latency prediction} We require an accurate predictor that provides a reliable per-request reward estimate which is a function of highly complex multiple system layers including the specific LLM serving engine (such as vLLM or SGLang, the version of the engine, and its configuration), size of the LLM model, model architecture, network condition, and hardware. It also depends on the workload. The latency profile will keep changing dynamically based on the combination of all the above factors and more. Hence, the predictor model must be able to capture this complex latency behavior by learning online with the real-time observation and quickly adapt to it.

There is another fundamental challenge in achieving a high-accuracy latency predictor model due to a \emph{circular dependency} between the routing policy, cluster state, and the request latency profile. The decisions made by the deployed routing policy itself affect latency behavior---creating a feedback loop where the policy changes the distribution to be predicted, and vice versa (Figure~\ref{fig:circular_dependency} and \S\ref{subsec:motivation-online}).

\paragraph{2. Efficient and reliable cloud-scale routing system}
\sysname design introduces additional overhead on the request critical path: collecting real-time cluster state\footnote{In the remaining of this paper, cluster state indicates the collection of instance states in the cluster.}, preparing the model inputs, and running the complex routing logic for each request.
These steps must be highly efficient, as any added latency directly offsets the gains from improved routing decisions. 
Also, the routing system must meet the reliability expectations of gateway infrastructure, continuing to operate correctly despite transient failures and unexpected slowdowns such as impact from GC, compute/network contention, etc.

\paragraph{\sysname summary}
To solve these challenges, we build \sysname, a data-driven online-learned distributed LLM inference routing system. It learns a reward predictor online for any given cluster, workload, LLM engine and LLM model. We use $-$TTFT as the main reward in this paper\footnote{We will describe why TTFT is a reasonable reward for routing optimization in \S\ref{sec:routing-policy}.}.
The reward predictor is a neural network that takes the request features and current cluster state as input and, for each serving instance, predicts the reward \sysname would obtain by routing the request there.
To learn online, \sysname continuously collects data at runtime while it is serving requests with the current predictor model. It learns a more accurate reward predictor over time by adapting to the current cluster configurations, workload, and importantly, the aforementioned circular dependency between the current routing policy, cluster state and eventually the latency profile. The features are selected to encode cluster state (LLM-aware fine-grained load metrics, KV cache state, hardware utilization) and request properties (lengths and expected prefix KV cache hit). 

The primary contributions of this work are:

$\bullet$ We present the design of a new distributed inference routing system, \sysname, that is adaptable to the dynamic environment via online learning.
We implement \sysname in an efficient and robust manner, and it is easily deployable in cloud-native infrastructure (Kubernetes).
\sysname has been built on top of \aibrix and is fully open-sourced\footnote{Code available at https://github.com/gangmuk/Lodestar}.

$\bullet$ We comprehensively evaluate \sysname's performance in a public cloud cluster with three cluster configurations, including both homogeneous and heterogeneous GPUs, with a variety of realistic workloads. 
Compared to the state-of-the-art LLM-aware routing policy (\heuristic), \sysname reduces mean TTFT by 1.41$\times$ overall and reduces P99 TTFT by 1.47$\times$ overall. \sysname reduces mean TTFT by 1.02--2.15$\times$ and reduces P99 by 1.07--1.86$\times$ in the homogeneous cluster, and reduces mean TTFT by 1.25--4.38$\times$ and reduces P99 by 1.32--4.42$\times$ in the heterogeneous cluster. 
This performance is achieved through roughly $\approx$5 minutes of online learning (\S\ref{sec:eval}).

$\bullet$ We show and analyze what the ``circular dependency'' effect is in a learning-based request routing system, why it is important to be aware of for learned policy, and how it manifests as a form of intertwined relationship between learned policy, induced cluster state, and request latency profile (\S\ref{sec:eval_adaptability}).

$\bullet$ Alongside the code, we release the first public per-request LLM routing dataset pairing cluster snapshots, request features, and latency collected from public-cloud clusters.

Modern clusters are becoming increasingly heterogeneous over time~\cite{gavel, mlaas_wild, helix, hetis} with new-generation accelerators~\cite{nvidia_blackwell, nvidia_rubin, google_tpu_ironwood, amd_mi300a}. A very active community is also quickly optimizing and evolving LLM serving engines. The LLM system stack is becoming even more distributed with memory tiering for KV cache~\cite{lmcache, mooncake, impress} and multi-cluster deployment. Given these ongoing changes, we expect that handwritten heuristics for inference routing will become even less practical, and data-driven, ML-based inference routing will bring even more performance benefits in the future.


\section{Background}
\label{sec:bg}
\paragraph{LLM inference serving} Unlike traditional web services, LLM requests exhibit extreme heterogeneity in computational cost, driven by widely varying input and output token lengths. Transformer-based inference has two phases with very different resource profiles. During \emph{prefill}, the model processes all input tokens in parallel and emits the first output token; this phase is compute-bound. \emph{Decoding} then generates subsequent tokens one at a time, doing little compute per step but repeatedly reading and writing the KV blocks of all previous tokens, making it memory-bandwidth-bound. Modern serving engines target these phases with optimizations such as prefix caching, continuous batching~\cite{orca}, chunked prefill~\cite{sarathi}, and speculative decoding~\cite{speculative_decoding}. Prefix caching in particular reduces time-to-first-token (TTFT) by caching the KV tensors of shared input prefixes and skipping their recomputation; because transformer attention is causal, prefix reuse is strictly sequential---cached blocks help only when all preceding blocks are also present. Note that token scheduling within a serving-engine instance and request routing at the gateway layer are distinct problems; \sysname is a request router, not a token scheduler.

\paragraph{Gateway infrastructure}
In LLM inference infrastructure, a gateway sits between clients and a pool of LLM engine instances, each running on its own GPU(s). This infrastructure offers LLM-specific routing policies that exploit \textit{prefix cache hit ratio}: the fraction of a request's input tokens whose KV blocks are already cached on a given instance. To compute this ratio, the gateway needs to track which prefix KV blocks each instance has cached. 
Each engine maintains its own local KV cache (typically with LRU eviction). To leverage it for smart routing in the gateway, the gateway maintains a prefix cache index---implemented as radix trees~\cite{sglang} or hash tables~\cite{aibrix}---that tracks which prefixes reside on which instances. The workflow is: the gateway tokenizes the incoming prompt, computes prefix matching ratio for the given prompt against each instance, runs a configured routing logic, routes the request to the selected instance, and updates the prefix tracking data structure.

\section{Motivation and Technical Challenges}
\label{sec:motivation}
In this section, we describe existing approaches and their limitations, motivate learning-based approach for LLM inference routing, and outline the key technical challenges.

\subsection{Existing LLM routing policies and their limitations}
\label{sec:existing_routing_policies}


\paragraph{Model-based approach} 
One category estimates each instance's expected latency analytically and routes to the instance with the lowest estimate. Mooncake/LMDeploy~\cite{mooncake, lmdeploy} approximates expected latency as queue length divided by throughput, treating throughput as a static constant and collapsing the state-dependent dynamics of LLM inference into a single fixed parameter. Preble~\cite{preble} models per-layer latency from offline profiles. These approaches share a fundamental problem. First, such LLM layer-wise latency models are likely to be inaccurate: request latency emerges from the full engine pipeline---queueing, batching and chunking, KV-cache pressure and eviction, memory-bandwidth contention across concurrent requests---not from a closed-form expression the designer can write down. Second, they do not scale: staying accurate would require re-profiling every combination of LLM model, data type, serving engine (vLLM, SGLang, etc.) and its version and configuration, and GPU---a non-separable cross-product that must be profiled jointly.


\paragraph{Rule-based approach} The other category follows a rule-based path: combine LLM-specific proxy signals typically prefix-cache hit benefit and queue state through threshold-gated or weighted rules to trade cache affinity against load. This is the most widely used approach and appears in all major LLM inference infrastructures: \sglang's \textit{cache-aware}~\cite{sglang_prefix_hash}, Dynamo's \textit{KV-aware routing}~\cite{dynamo_kv_routing}, llm-d's \textit{precise-prefix-cache-scorer}~\cite{llmd_routing}, and \aibrix's \textit{prefix-cache-and-load}~\cite{aibrix_prefix_load_routing}. They differ in form but share a deeper limitation: any fixed rule over a small set of hand-picked signals cannot express how latency actually depends on cluster state, which is nonlinear and joint across many more factors (we expand on this in \S\ref{subsec:motivation-why-ml}). In the rest of the paper we use \heuristic (Appendix~\ref{sec:appendix_alg}, Algorithm~\ref{alg:prefix_and_load_aware_routing}), \aibrix's \textit{prefix-cache-and-load} routing as our primary heuristic baseline.

\begin{figure}[!t]
    \center
    \begin{subfigure}[b]{0.95\columnwidth}
        \includegraphics[width=\textwidth]{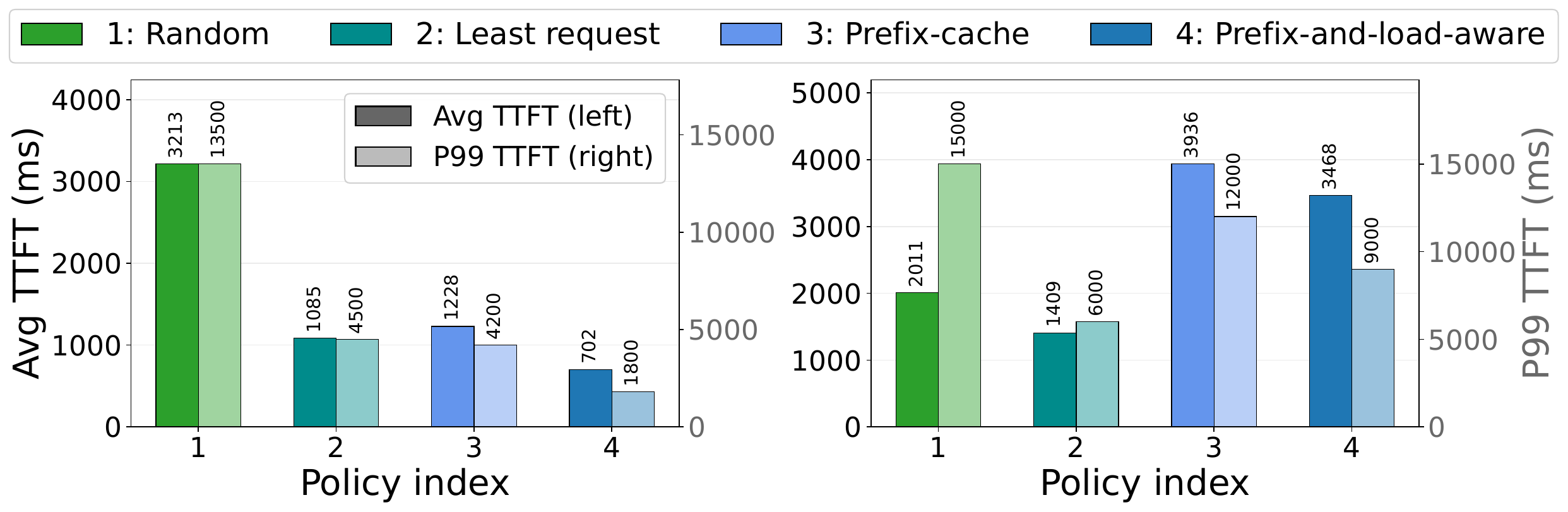}
    \end{subfigure}
    \caption{TTFT performance of the four existing routing policies under a high prefix sharing (80\%) workload. Experiments were conducted on a seven-NVIDIA-L20 GPU instance cluster using the DeepSeek 7B model with FP16 precision. Two different workloads were used for each experiment (Left: RPS=5, avg input length=4K, Right: RPS=10, avg input length=1K).}
    \label{fig:motivation_existing_routing_polices_perf}
\end{figure}

\begin{figure}[!t]
    \centering
    \begin{minipage}{0.5\columnwidth}
        \includegraphics[width=\textwidth]{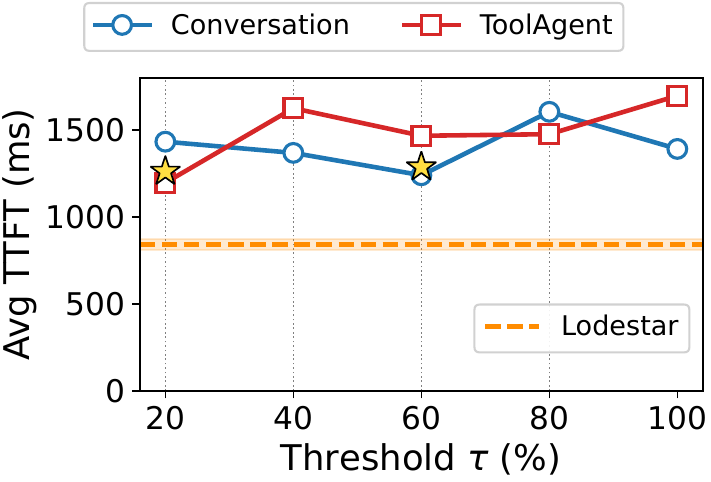}
    \end{minipage}
    \hfill
    \begin{minipage}{0.47\columnwidth}
        \caption{\prefixaware routing with different prefix hit thresholds. The threshold ($\tau$) means the minimum prefix hit ratio to enable prefix-aware routing. If the max possible prefix hit ratio across all instances is lower than the $\tau$, it uses least request.}
        \vspace{-0.2in}
        \label{fig:motivation_threshold_tuning}
    \end{minipage}
\end{figure}



\subsection{The case for learned routing}
\label{subsec:motivation-why-ml}

Figure~\ref{fig:motivation_existing_routing_polices_perf} shows two workloads with the same 80\% prefix-sharing\footnote{Prefix sharing ratio: fraction of input tokens whose prefix is shared with other requests in the workload.} but different requests per second (RPS) and input length. In the low-RPS, relatively longer-input regime (Fig.~\ref{fig:motivation_existing_routing_polices_perf}a), \heuristic achieves the lowest TTFT (702\,ms avg, 1800\,ms P99). In the high-RPS, relatively shorter-input regime (Fig.~\ref{fig:motivation_existing_routing_polices_perf}b), the same policy degrades $\sim\!5\times$ to 3468\,ms avg and 9000\,ms P99, and is beaten by Least-Request---a naive load balancer with no LLM-specific logic (1409\,ms avg, 6000\,ms P99).

One might hope a better-tuned threshold would close the gap. Figure~\ref{fig:motivation_threshold_tuning} sweeps the prefix-hit threshold $\tau$ of \prefixaware routing from 20\% to 100\%\footnote{at $\tau=100\%$ the policy effectively reduces to least-request load balancing}. The two workloads' optima sit at different thresholds---ToolAgent at $\tau=20\%$, Conversation at $\tau=60\%$---and even each workload's best threshold (stars) still sits $\sim$300--400\,ms above \sysname (dashed). 

The limitation of the existing heuristic approaches is fundamental. The mapping from cluster state to TTFT is nonlinear, depends jointly on input length, load, prefix hit ratio, and cluster configuration---factors that interact non-separably---and shifts with the workload and cluster setup itself. No fixed rule over a handful of hand-picked signals can represent such a mapping, which is why threshold tuning on a single signal cannot close the gap observed above. The routing policy should instead be learned from data. A neural network is a natural model class for this: it can fit nonlinear, high-dimensional functions directly from observed latency, capturing the joint dependence of latency on request, instance, and cluster signals that no fixed rule can express.

\begin{figure}[!t]
    \centering
    \begin{subfigure}[b]{0.4\columnwidth}
        \includegraphics[width=\textwidth]{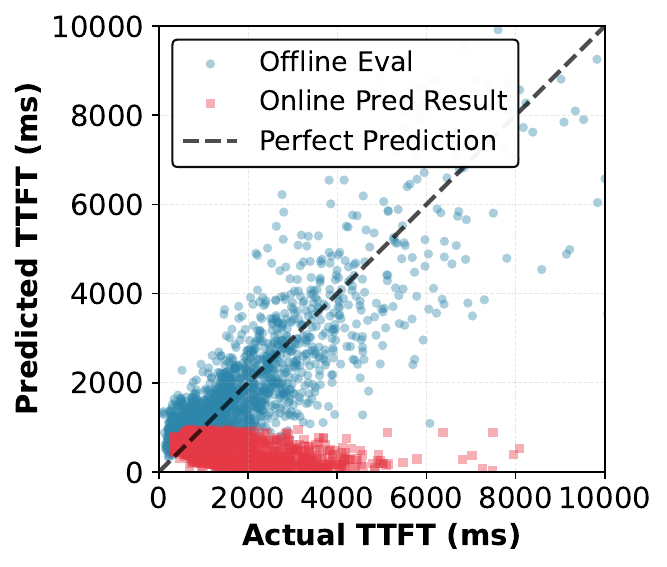}
        \caption{Effect of the circular dependency on prediction.}
        \label{fig:circular_dependency_scatter_plot}
    \end{subfigure}
    \hfill
    \begin{subfigure}[b]{0.59\columnwidth}
        \includegraphics[width=\textwidth]{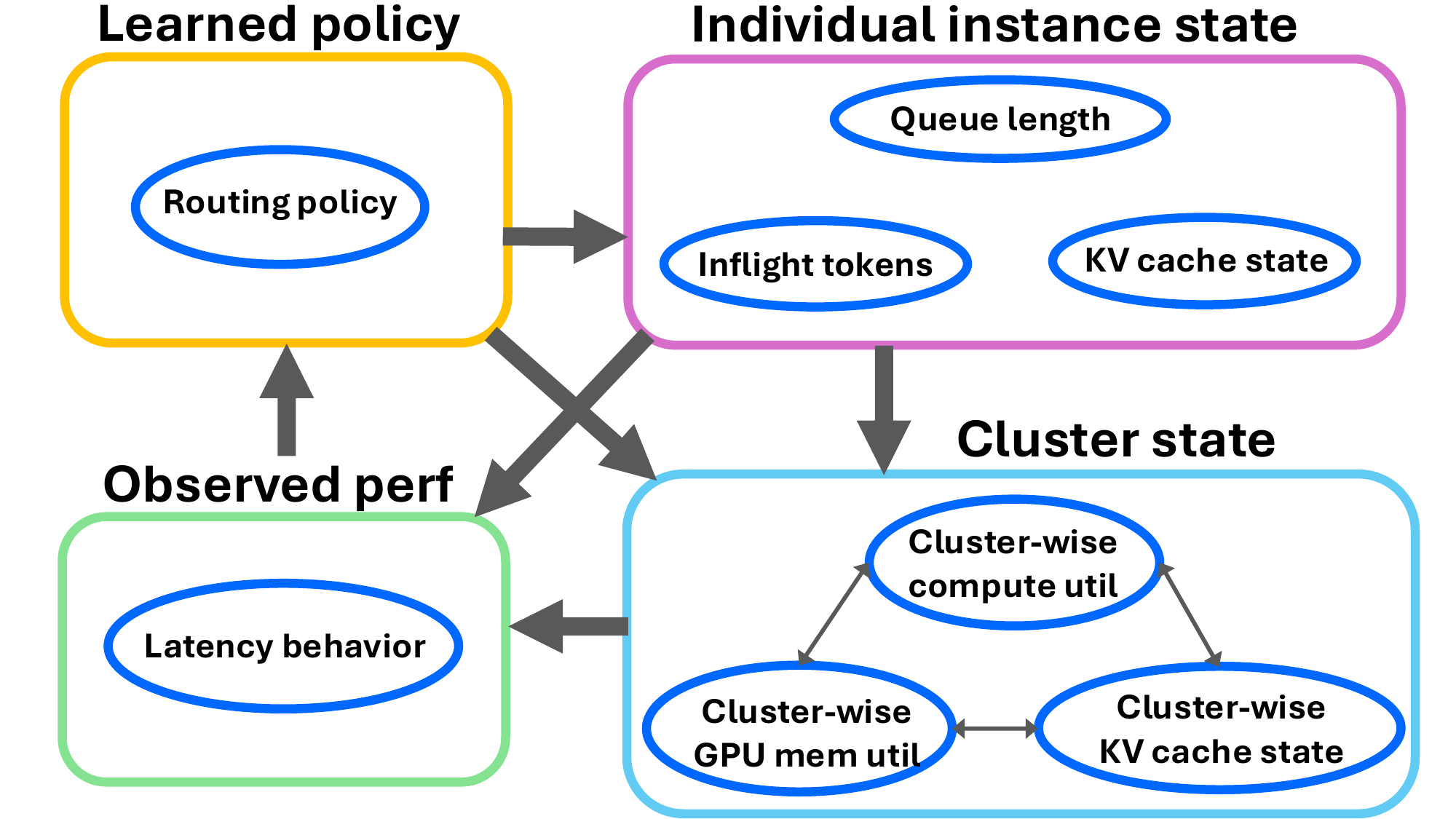}
        \caption{Circular dependency diagram.\\\strut}
        \label{fig:circular_dependency_diagram}
    \end{subfigure}
    \caption{Figure ~\ref{fig:circular_dependency_scatter_plot} shows the accuracy performance of the offline trained model when it is evaluated offline and when it was evaluated online. Blue dots show the offline predictions of the offline trained model. Red dots show the online predictions when the same offline trained model was used online. The diagonal line represents perfect prediction. The performance degradation highlights the necessity of an \textbf{online adaptive} ML model.
    Figure ~\ref{fig:circular_dependency_diagram} is a visual description of the circular dependency.
    }
    \label{fig:circular_dependency}
\end{figure}

\subsection{The case for online-learned routing}
\label{subsec:motivation-online}

ML models generalize well only when the training data matches the deployment distribution and covers its operating conditions with sufficient diversity. In LLM serving, both conditions are hard to meet offline. New workloads may not have been observed at training time---a coverage failure---and even for the same workload, the feature distribution shifts with cluster configuration, serving-engine state, and the routing policy in effect---a distribution-shift failure. It is therefore impractical to assume that offline data covers all the cases a deployed predictor will encounter.

Even under the optimistic assumption that offline data is sufficiently comprehensive and distribution shift is negligible, there is a second and more fundamental reason a learned routing policy must learn online: \emph{circular dependency}.\footnote{The ML community refers to a closely related phenomenon as \emph{compounding error} in imitation learning~\cite{dagger11ross}.} The routing policy and the data distribution it observes are mutually determined. The policy learns from past data, updates itself, and by doing so changes the cluster state that future requests encounter; the changed state in turn changes the latency profile, which changes the next predictor's training data. A predictor trained under one policy and deployed to drive a different policy is never evaluated on the distribution it will actually face.

Figure~\ref{fig:circular_dependency_scatter_plot} illustrates this in practice. We trained a TTFT predictor offline on data collected from \S\ref{subsec:motivation-why-ml}'s workloads using Least-Request and \heuristic routing (since no accurate predictor exists before the first learned policy is trained). On held-out offline data, the model's prediction is relatively accurate (blue dots) by clustering around the diagonal. However, when the same model is deployed to route requests, its predictions collapse (red dots). The model is systematically over-optimistic and loses all ranking signal between instances---fatal for a routing rule that picks $\arg\min$ over predicted TTFT (equivalently, $\arg\max$ over predicted reward). Offline training alone is insufficient; online adaptation is necessary.

\sysname therefore learns \emph{online}: data is collected continuously during deployment, and the predictor is retrained periodically so that each new model trains on data drawn from the distribution its predecessor just induced. Online learning addresses distribution shift by construction, and while circular dependency persists, each retraining round narrows the gap between the predictor's training distribution and the current operating distribution. This demands that the online-learning algorithm be both \emph{accurate} enough to produce useful routing decisions and \emph{quickly adaptive} enough to close the gap before the distribution moves further.

\subsection{ML on online LLM serving critical path}
\label{subsec:motivation-critical-path}
Request routing sits on the critical path, so any routing-side computation must be budgeted against request latency.
TTFT spans hundreds of milliseconds to seconds, so running more complex routing logic with additional overhead for LLM inference can be compensated with much larger margin. And the per-decision value is also much higher: each decision commits an expensive GPU for a relatively long time and, through shared KV cache and queue state, propagates to subsequent requests---so a single bad choice can waste seconds of accelerator time. This newly widened budget is what makes a rich-feature, neural-network-based router affordable for LLM serving; we quantify the resulting overhead in \S\ref{sec:eval_overhead}.


\begin{figure}[t]
    \centering
    \begin{subfigure}{0.48\textwidth}
        \centering
        \includegraphics[width=\textwidth]{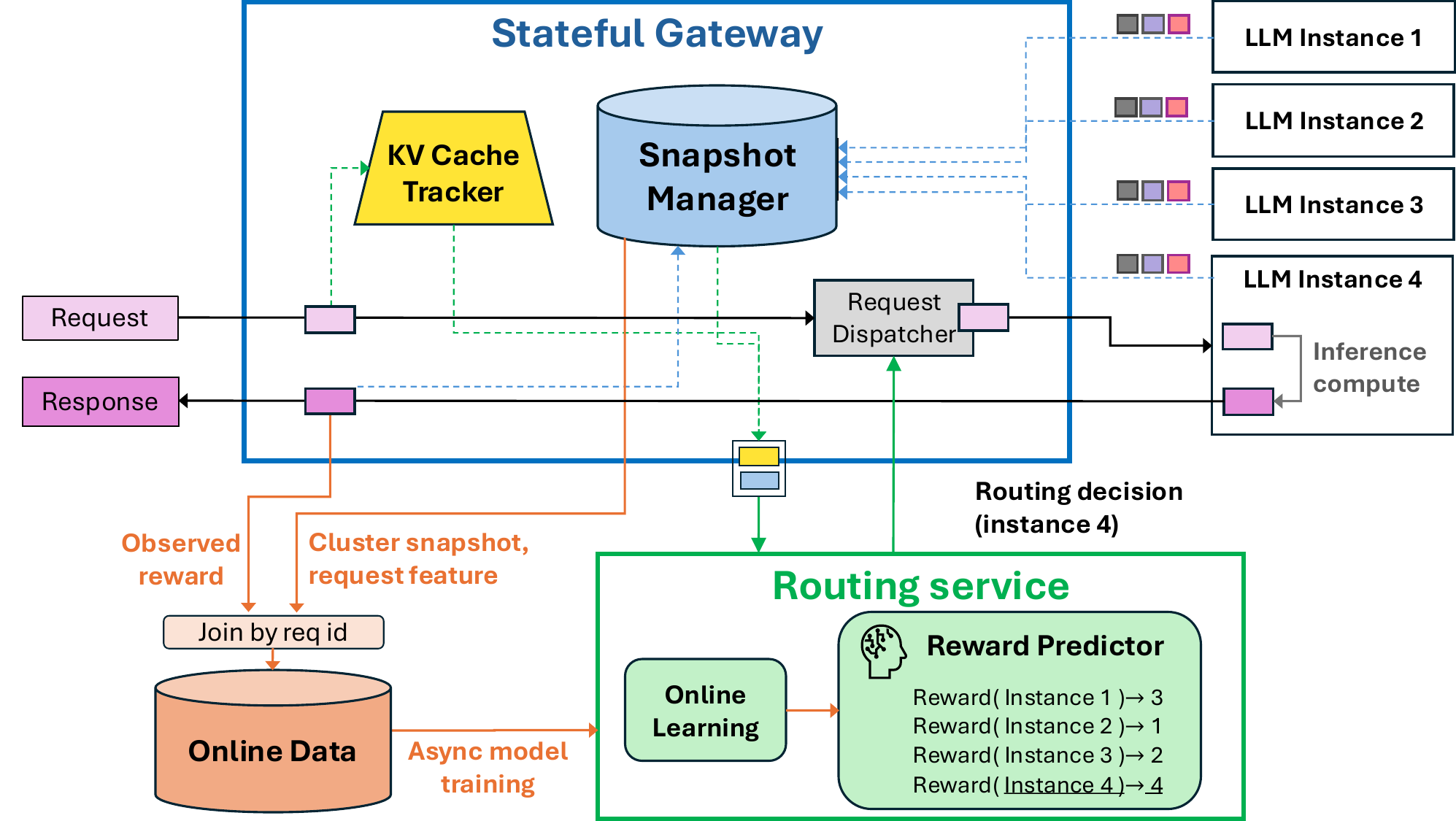}
        \caption{\sysname architecture.}
        \label{fig:arch}
    \end{subfigure}
    
    \vspace{0.5em}
    
    \begin{subfigure}{0.48\textwidth}
        \centering
        \includegraphics[width=\textwidth]{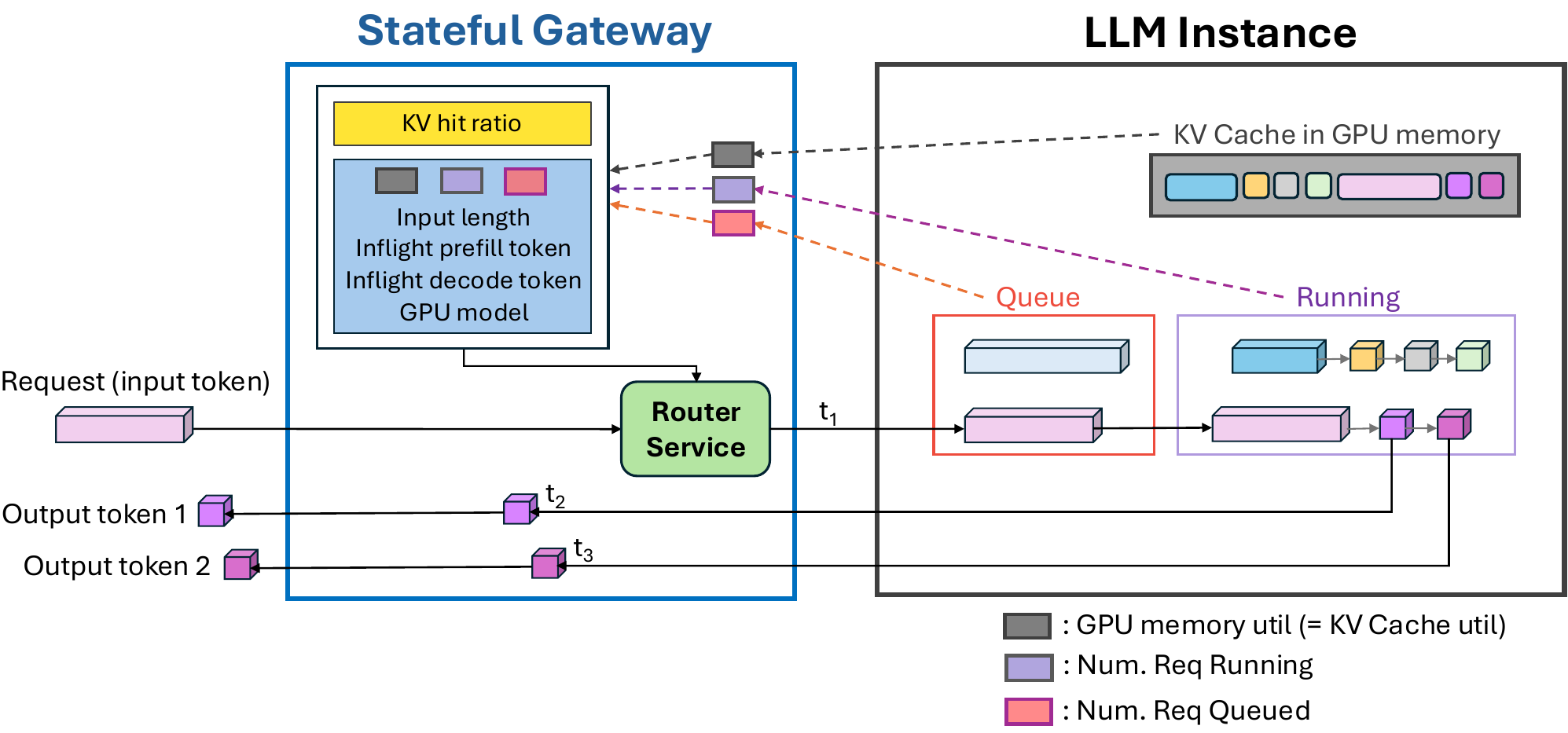}
        \caption{Data collection pipeline architecture.}
        \label{fig:data_collection_arch}
    \end{subfigure}
    \caption{Overall architecture.}
    \vspace{-0.15in}
    \label{fig:combined_arch}
\end{figure}


\section{\sysname Design}
\label{sec:design}
\sysname's design (Figure~\ref{fig:arch}) consists of two primary components: the \gatewayfull and \ras. The whole system is designed to enable data-driven request routing decisions based on request features and cluster state. On the request path, \textit{\gatewayfull (\S\ref{sec:design_gateway})} takes a cluster snapshot at every request arrival, passes it to the \ras for routing computation. On the response path, it records the observed latency at per-token level. All data is indexed by request ID in an online database and used for online learning.
\textit{\ras (\S\ref{sec:design_ras})} runs the reward predictor model inference for routing decision given the current cluster snapshot provided by \gatewayfull. For online learning (\S\ref{sec:design_online_learning}), \ras asynchronously updates the reward predictor model with data collected by \gatewayfull every time there is $\theta$ number of new samples. 1000 was used for $\theta$ in the evaluation. 
\gatewayfull and \ras run separately in a distributed way in the cloud infrastructure to isolate from potential failure, slowdown, or unreliable decision of \ras. 

Next, we describe these components in more detail, beginning with the learning-based routing policy (\S\ref{sec:routing-policy}), which shows how \sysname makes decisions for each request. We then cover the system design of \gatewayfull (\S\ref{sec:design_gateway}) and \ras including the online learning mechanisms (\S\ref{sec:design_ras}), and finally describe our implementation (\S\ref{sec:implementation}).

\subsection{Learning-based routing}
\label{sec:routing-policy}

We model a scalar \emph{reward} for each (instance, request) pair, and the request is routed to the instance with the highest predicted reward.


\paragraph{Request-instance reward model}
We define the reward as the negative of observed TTFT\footnote{TTFT is reported in seconds.}, $y_{i_{max}} = -\text{TTFT}_{i_{max}}$, where $i_{max}$ is the selected instance index. 
Let $\mathbf{x}_i$ denote the feature vector for instance~$i$ and $\mathbf{r}$ the request feature vector for the current request. A single \emph{reward model} $f_\theta$ scores each instance independently as $\hat{y}_i = f_\theta([\mathbf{x}_i \,\|\, \mathbf{r}])$ for $i = 1,\ldots,N$, where $[\cdot \| \cdot]$ denotes concatenation and $N$ is the number of active instances. The request is routed to instance $\arg\max_i \hat{y}_i$, i.e., the instance with the highest predicted reward. 
Our approach is thus a prediction-driven greedy routing strategy. 
\sysname optimizes time-to-first-token (TTFT) latency for two reasons: TTFT is critical to the user experience, and Time-per-output-token (TPOT) or end-to-end latency is not the right metric to optimize. The time that TPOT and end-to-end latency are observed is much further in the future than the routing decision time. Additionally, they are affected by one more unknown factor, the number of decode tokens, becoming even a more noisy signal.

\begin{figure}[!t]
    \centering
    \includegraphics[width=0.8\columnwidth]{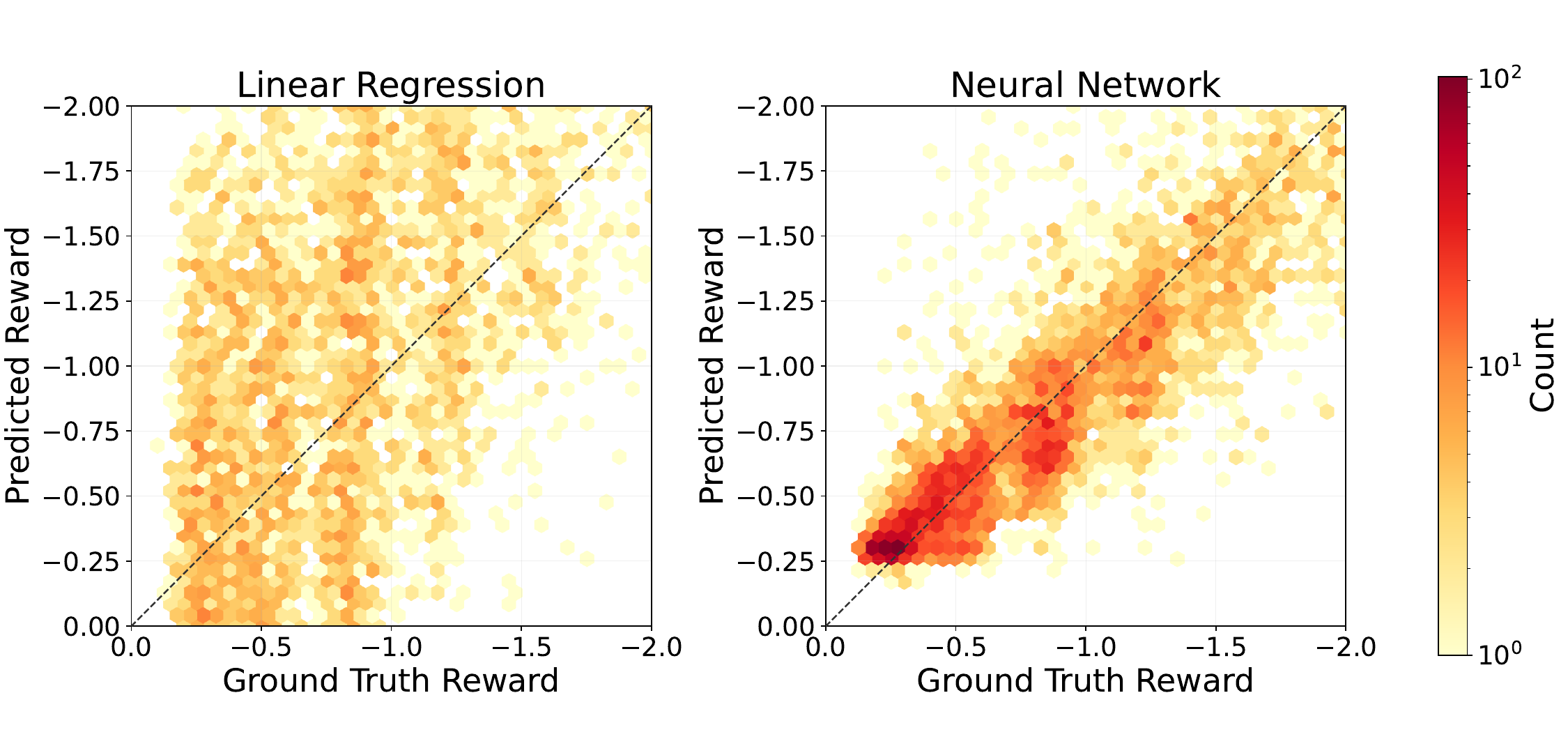}
    \caption{Reward (TTFT) prediction performance in linear regression and neural network.}
    \vspace{-0.1in}
    \label{fig:linear_regression_vs_neural_network}
\end{figure}

\paragraph{Model architecture} The reward model $f_\theta$ is an MLP with 3 hidden layers of 128 units and a scalar output, using ReLU activations and dropout (0.1) between hidden layers. Crucially, the same parameters~$\theta$ are shared across all instances, and instance identity is never an input. This has two consequences. First, the architecture is \emph{instance-count independent}---it works unchanged as instances are added or removed, which is essential for elastic cloud deployments. Second, it is \emph{instance-index independent}, which avoids a ``herding effect'' in which the model would associate a particular instance ID with good performance, over-route traffic to it, and then oscillate between overloading and underloading it across training rounds. 
To show why we need neural network instead of a simpler ML model such as linear regression, Figure~\ref{fig:linear_regression_vs_neural_network} compares linear regression against the \sysname predictor, using identical features and training/evaluation data. The neural network's predictions cluster close to the ground-truth diagonal, while linear regression exhibits larger errors and higher variance. The gap reflects that the mapping from features to latency is fundamentally non-linear (\S\ref{subsec:motivation-why-ml}).

\paragraph{Features}
Choice of features is very important to make the reward predictor accurate. To empower the predictor, we use a rich set of features from three different sources: request, expected KV cache hit ratio, and instance state. 

\noindent(1) \textit{Request feature} includes input token length. Input token length is directly related to the amount of prefill computation. Also, it affects how many new KV blocks would be allocated for this request. Both are affected by combination of input token length and KV cache hit ratio.

\noindent(2) \textit{Expected KV cache hit ratio} is obtained by the KV cache tracker in \gatewayfull. The input prompt is tokenized and inserted into the prefix index data structure. For a given input prompt of an incoming request, each instance will have a different expected KV cache hit ratio depending on what requests it served in the past. It tells how much portion of prefill computation can be saved when the request is routed to the instance. Also, it affects how many new KV blocks would be needed for the request.
It does not belong to the instance state since KV cache hit has not happened. This feature is the same one that existing heuristics use in their rules. 

\noindent(3) \textit{Instance state}. Two per-request load metrics: the number of running requests (compute load) and the number of queued requests (queueing delay). Two per-token load metrics: inflight prefill tokens and inflight decode tokens. We keep prefill and decode tokens separate because prefill is compute-bound and decode is memory-bandwidth-bound. This implies the effective utilization of compute and memory bandwidth resources.
Collapsing them would hide which bottleneck the instance is approaching. For hardware state, we include GPU memory utilization, which directly affects request schedulability and preemption, and the GPU model name as a categorical feature to support heterogeneous clusters.

These features are not separable, rather having an impact on each other. The above description is not necessarily the only implications of them. 

\noindent\textit{Exclusions.} A reader might expect instance state to include common hardware-utilization metrics\footnote{GPU utilization is exposed by NVML, and DCGM. SM means Streaming Multiprocessor: compute unit in GPU. GEMM means GEneral Matrix Multiply. Fine-grained GPU profilers are Nsight Systems and CUPTI.}---GPU utilization, SM activity, and memory-bandwidth utilization. We deliberately leave them out. GPU utilization is misleading by name. It reports 100\% whenever any kernel is running on any SM, regardless of how many SMs are actually working or how hard. A tiny kernel on one SM reads the same as a kernel saturating the entire GPU. SM activity and memory-bandwidth utilization are finer-granularity metrics, but we still do not use them as features. Each monitoring sample averages the resource usage of running GPU kernels in its window. Different sampling windows catch different sets: one window's set might be mostly compute-bound prefill GEMMs that saturate the SMs, the next mostly memory-bandwidth-bound decode attention that saturates memory bandwidth, depending on which requests were in which phase at the time. The reading bounces around even when the underlying load is steady, and the predictor cannot separate real load changes from this sampling artifact. Fine-grained profilers can sample at nanosecond resolution, but their overhead is incompatible with online routing. In conclusion, these metrics would add more noise than signal, hurting model accuracy and slowing convergence. 

\paragraph{Consistent hashing based filtering}
\sysname's selection of the instance with maximum predicted reward is greedy.  Even though a greedy decision is not necessarily clearly suboptimal, by design it cannot consider decisions' long-term global effect. For example, even if an incoming request $R$ is routed to the instance that is best (i.e., produces minimum latency) for $R$, it can cause evictions of KV objects from the cache that could have been hit by other future requests, and this eviction might be globally suboptimal. We found such a situation occurs when the GPU memory resources (KV cache utilization) in the entire cluster is saturated and there are many requests that share long prefixes. Hence, the eviction of that KV has big impact globally across many requests.

To overcome the limitation of greediness, we introduced \emph{Consistent hashing based filtering}. The basic idea is to guide the system to concentrate each shared prefix's KV cache state on a smaller number of servers, while still providing enough options for the reward predictor to be effective.  Specifically, when the cluster GPU memory is saturated (cluster memory utilization > X\% \footnote{We used 80\% but the optimal config will vary depending on the cluster and workload.}), \sysname first filters the set of all instances by applying consistent hashing with $k$ hash functions, with the shared prefix group as the hash key, to select $k$ instances. 
Within those $k$\footnote{We used 2 in our evaluation.}, we then greedily select the instance with maximum predicted reward. This nudges the routing of requests with shared prefixes onto a tighter set of instances, making better use of limited global cache space.

We show in \S\ref{sec:eval_k_filtering} that Consistent hashing based filtering is effective, but we do not claim it is the only or best approach to moving the system towards the global optimum. Our current solution has some clear limitations in tuning the parameters of cluster saturation and $k$. We leave further explorations of this direction to future work.

\subsection{\gatewayfull}
\label{sec:design_gateway}
The \gatewayfull receives all inference requests and coordinates routing across LLM instances. For each incoming request, it collects real-time metrics from all active instances and maintains a prefix index to estimate the expected KV cache hit ratio. These signals are assembled into a feature vector that captures the current system state and is passed to the \ras. As a fallback, the \gatewayfull runs a heuristic that takes over when the \ras is unavailable or fails. A complete routing algorithm on \gatewayfull is listed in Appendix~\ref{sec:appendix_qs} Algorithm~\ref{alg:routing-with-fallback}.

\paragraph{Cluster snapshot manager} 
Embedded within the \gatewayfull, the \textit{Cluster snapshot manager} takes a snapshot of the cluster (all instances' state) at every request arrival. The requirement is low overhead since it is in the request serving critical path. 
Cluster state is composed of two feature groups, one that can only be learned from application internal state and the other that can be tracked by \gatewayfull. 
The first group includes GPU memory utilization (KV cache util), internal waiting queue state, and internal running queue. These are scraped proactively from the LLM serving instances every $X$ milliseconds\footnote{100ms was used in our setup.} on background, so they do not affect the critical path (small gray, purple, red boxes in Figure~\ref{fig:data_collection_arch}). 
Per-token load metrics (inflight prefill/decode tokens) are tracked solely by \gatewayfull without interacting with instances as it sees all the token-level interaction. Also, input length and expected prefix hit are computed by the \gatewayfull at the request arrival time. GPU model feature is updated only when an instance is added to the cluster. 

Minimizing overhead of state collection requires careful implementation. The snapshot manager does not have any lock between different requests and between different features in cluster state data structure. Overhead is discussed further in \S\ref{sec:design_requirements_ras} (P1).

Cluster state data are collected in an in-memory dictionary with request ID as the key and will be periodically flushed to the \ras on background to be persisted for online training. The metrics it collects are detailed in \S\ref{sec:routing-policy}.

\paragraph{Prefix KV cache tracker}
The \gatewayfull keeps track of prefix placement based on its own routing history and runs prefix matching logic to calculate expected prefix KV cache hit ratios. We use a logical radix tree data structure as commonly used in modern LLM routing stack (\S\ref{sec:bg}). The input is a tokenized input message of a request, and the output is the expected prefix KV cache hit ratio for each instance. 
This is similar to how serving engine instances such as vLLM or SGLang maintain their prefix caches internally, but \sysname's routing infrastructure needs to track the global view of it to tell how much prefix would be shared at each pod for each incoming request.

\subsection{\ras}
\label{sec:design_ras}


\ras has three main roles: running the routing logic, persisting the cluster snapshots, and continuously updating the predictor model online.

\subsubsection{Running the routing logic}
\label{sec:design_requirements_ras}

The Routing Service (\ras) owns the learned routing logic: predicting rewards and training the predictor.
We described the routing model in \S\ref{sec:routing-policy}. Here we describe the system design surrounding that logic.
Its design follows three design requirements that answer the challenges in \S\ref{sec:motivation}: (P1) inference on the critical path must be fast, (P2) training must not stall inference, and (P3) \gatewayfull availability must not depend on \ras. 

\paragraph{P1: Bounded critical-path inference}
Every request pays the cost of one \ras forward pass, which must stay small. The main scaling concern is instance count $N$: scoring $N$ candidates naively requires $N$ forward calls\footnote{It includes paying CPU$\to$GPU transfer, Python/CUDA launch, and feature tensor construction each time}. To minimize overhead, \ras batches features of $N$ candidate instances into a single input of shape $[N, d]$ ($d$ is the per-instance feature dimension) and performs one forward pass, amortizing per-call overhead. If $N$ grew large enough that even this mattered, \sysname could simply use the standard power-of-$d$-choices technique (for some $d \ll N$) which has been widely adopted by other schedulers and load balancers~\cite{ptwo01mitzenmacher}.

Several other implementation points were important for \ras's efficiency. We had to avoid using any expensive library like Pandas, which is popular for data processing in ML but introduces a few milliseconds to hundreds of milliseconds of overhead while creating objects (DataFrame) and processing data. Also, \ras should only use $O(1)$ read/write data structures. We found these small things mattered to avoid overhead that would have made \sysname impractical.
Additionally, the \ras instance is scheduled in a GPU instance for fast inference compute. 
Together these hold per-request \ras overhead to a few milliseconds (Fig.~\ref{fig:overhead_analysis}), which is much smaller than the TTFT reduction achieved by \sysname.

\paragraph{P2: Inference/training isolation}
Training requires more compute and it must not interfere with millisecond-scale inference routing logic execution. Inference and training run as separate asynchronous tasks in \ras; when a newly trained model is ready, \ras atomically swaps the model pointer from the current model to the new one with zero downtime. 

\paragraph{P3: Fail-safe boundary with \gatewayfull} The \gatewayfull must remain available if \ras fails. \ras runs as a separate service, isolating failures from the gateway data plane; every \gatewayfull$\to$\ras RPC has a short timeout for transient slowness (CPU contention, network jitter, GC pauses), and the \gatewayfull pre-computes a heuristic selection before issuing the RPC, so on timeout or error the pre-computed choice is used immediately without added latency.

\paragraph{Fallback for efficiency and reliability}
\label{sec:design_fallback}
Three conditions trigger fallback to the heuristic: (i) \textit{cold start} --- the predictor has not yet been trained, or a newly loaded checkpoint's normalization statistics do not match current data; (ii) \textit{out-of-distribution inputs}, detected by a per-sample check that the request's features and the per-instance features lie within the ranges observed in the training buffer; and (iii) \textit{timeout or failure}, when the \gatewayfull$\to$\ras RPC times out or errors, handled by the pre-computed-heuristic mechanism described in P3 so the fallback adds no latency. 
(iii) is detected at the \gatewayfull side.
Note also that in all above cases the fallback is temporary because online learning will run with newly observed data. Approaches like this, often called ``guardrails'', are a research problem in their own right in learning-based systems~\cite{guardrail}. Adopting a more advanced guardrail algorithm could further improve the reliability of \sysname. We list the routing logic in Appendix~\ref{sec:appendix_qs} Algorithm~\ref{alg:ras-infer}.

\subsubsection{Online learning}
\label{sec:design_online_learning}
The online learning loop raises two system questions: managing the training data collection pipeline efficiently, and deciding which of that data the predictor trains on. 

\paragraph{Training-data collection}
Training data flows from \gatewayfull to \ras in batched, asynchronous flushes---per-request records are accumulated in \gatewayfull in memory and pushed in bulk (currently it is 100 requests per batch), amortizing I/O and keeping the request path untouched. The pipeline is best-effort: an occasional dropped batch only delays the next training round, since retraining is periodic and tolerant of missing samples. 

\paragraph{Training-data selection}

The choice of training data has the largest impact on predictor quality. For the predictor to be accurate online, training data must satisfy two properties. It must have the \textbf{recency} property, giving the predictor \textbf{adaptability} to the current workload and cluster state. It must also have the \textbf{diversity} property, because a model can only learn the effect of a feature when that feature varies in the training data, and predictions on inputs outside the training distribution's support are extrapolations with no accuracy guarantee. Training data concentrated in a narrow regime therefore yields a predictor accurate only within that regime---a limitation that matters for \ras, since the cluster continually shifts between regimes and drifts into new ones.

Neither of the two obvious schemes achieves both properties. A FIFO sliding window preserves recency but loses diversity: a stable cluster fills the window with homogeneous samples, and a workload shift floods it with the new regime, causing the model to forget earlier ones. Training on the full history preserves diversity but loses recency, blurring the model across stale distributions and growing training time unboundedly.

\ras achieves both with a \textit{two-pool} design. A \textbf{FIFO buffer} of size $|\mathcal{F}| {=} 5000$ holds the most recent samples, evicting the oldest on overflow. A \textbf{replay buffer} of size $|\mathcal{R}| {=} 5000$ begins to accumulate once the FIFO is full; samples evicted from the FIFO are admitted only if selected by a gradient-coreset criterion~\cite{gcr22tiwari}---samples whose last-hidden-layer activations, weighted by prediction residual, are most diverse with respect to those already kept. This keeps the replay buffer \emph{informative} rather than merely \emph{old}, covering regimes the model still mispredicts. Each training round uses $\mathcal{F} \cup \mathcal{R}$, and total storage is capped at $|\mathcal{F}| + |\mathcal{R}|$. We evaluate these design choices in Figure~\ref{fig:training_data_selection_algorithm_ablation_study}.


\subsection{Implementation}
\label{sec:implementation}
We implemented \sysname on top of \aibrix~\cite{aibrix}, a widely used open-source LLM inference infrastructure. The \gatewayfull is implemented in 8K lines of code in Go, and \ras is implemented in 4K lines of code in Python. It is designed to be cloud-native, and all components runs in K8S clusters. The full implementation is available at the repository linked in \S\ref{sec:intro}.

\begin{figure*}[!t]
    \centering
    \includegraphics[width=0.9\textwidth]{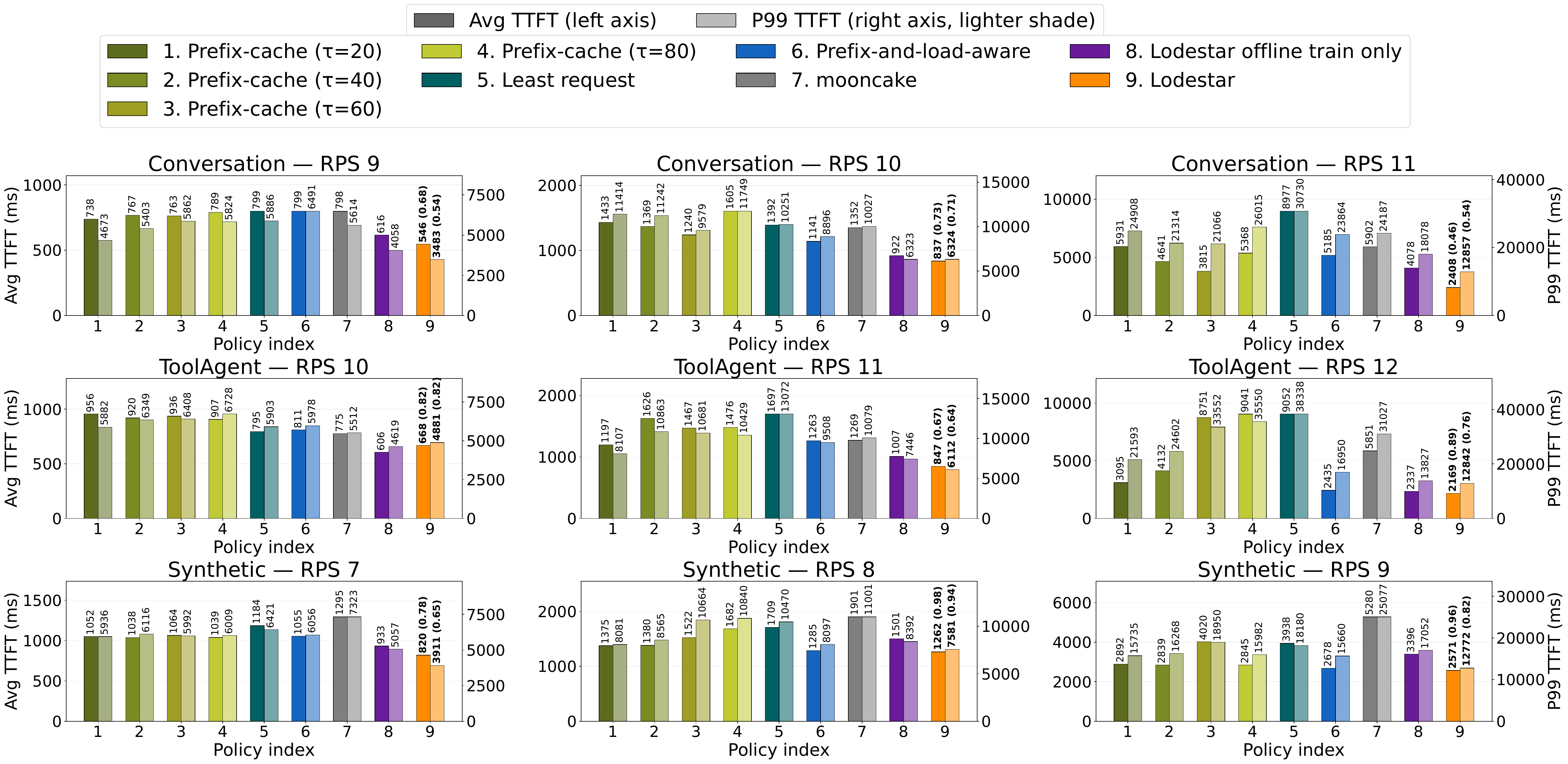}
    \caption{TTFT latency comparison for Mooncake workloads in aggregated deployment. The numbers noted inside the parentheses above \sysname are its relative performance against \heuristic.}
    \label{fig:ttft_mooncake_homo_aggregated}
\end{figure*}

\begin{figure*}[!t]
    \centering
    \includegraphics[width=0.8\textwidth]{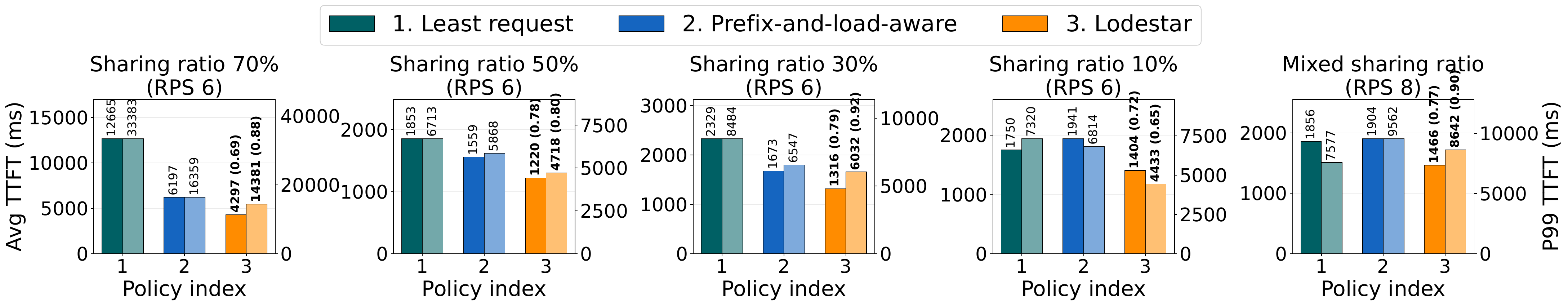}
    \caption{TTFT latency comparison for different prefix sharing ratio workloads (10\%, 30\%, 50\%, 70\%, Mixed\%) in aggregated deployment.}
    \label{fig:ttft_gangmukprefix_homo_aggregated}
\end{figure*}

\begin{figure}[!t]
    \centering
    \includegraphics[width=0.49\textwidth]{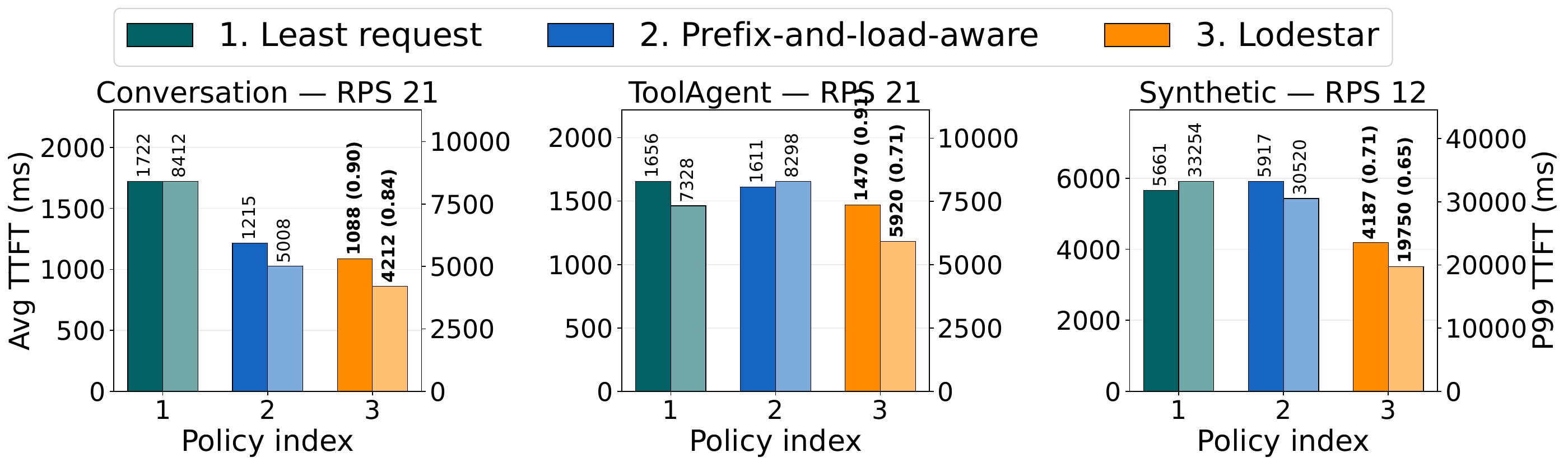}
    \caption{TTFT latency comparison for prefill-only workload.}
    \label{fig:ttft_mooncake_homo_prefillonly}
\end{figure}


\section{Evaluation}
\label{sec:eval}

In the evaluation, we answer the following questions:

\noindent 1. How does \sysname's routing compare against baselines across workloads on homogeneous (\S\ref{sec:eval_homo_cluster}) and heterogeneous (\S\ref{sec:eval_hetero_cluster}) clusters?

\noindent 2. Can \sysname's online learning adapt to workload shifts, and how does circular dependency manifest in the learned routing policy (\S\ref{sec:eval_adaptability})?

\noindent 3. Does \sysname achieve low overhead in its advanced routing decision pipeline (\S\ref{sec:eval_overhead})?

\noindent 4. Does \sysname's online training data selection strategy work well (\S\ref{sec:eval_data_selection_strategy})?


\subsection{Experimental setup}


\paragraph{Cluster setup}
We tested \sysname in two different cluster setups: a homogeneous GPU cluster with eight NVIDIA-A30 GPUs (24GB GPU memory) and a heterogeneous GPU cluster with eight NVIDIA-A30 GPUs and eight NVIDIA-V100 GPUs (32GB GPU memory). Each NVIDIA-A30 instance has Intel Xeon Platinum 8336C @ 2.30GHz CPU (28 cores) and 240 GB host memory. Each NVIDIA-V100 instance has Intel Xeon Platinum 8260 @ 2.40GHz (8 cores) and 32 GB host memory. All instances run in a public cloud Kubernetes cluster. One of the A30 instances is used for \ras.

\paragraph{Workloads}
We use the Mooncake data set~\cite{mooncake}  which has a conversation workload, a tool\&agent workload, and a synthetic workload. Conversation is from a chatbot application; toolagent is from an agentic application with tool calls; and synthetic is constructed by mixing three datasets: ShareGPT~\cite{sharegpt}, LeVal~\cite{leval} and LooGLE~\cite{loogle}. Figure~\ref{fig:mooncake_workloads} in Appendix~\ref{sec:appendix_exp} shows the distribution of the three workloads. 
It is important to note that the distribution of prefix-sharing ratios, prefix reuse distances within a shared-prefix group, and the number of requests per group all play important roles in LLM inference workloads.

To evaluate \sysname with a wider range of workload distributions, we synthesize five more workloads with different prefix sharing ratios based on publicly available statistics~\cite{alibabakvcacheworkload}. The first four workloads have 10\%, 30\%, 50\%, and 70\% average prefix-sharing ratios to represent different categories of workloads (single-turn chatbot, shared system prompt application, multi-modality, RAG, etc.). 
One additional workload mixes all four ratios (10\%+30\%+50\%+70\%) in equal proportion. Each workload contains multiple prefix groups, with requests within a group sharing the configured prefix ratio. Input lengths range from 1000 to 10000 tokens, and the output length is sampled from a normal distribution with mean=100 and std=10. Prefix reuse distance follows a uniform distribution.

The request inter-arrival time is sampled from a Poisson distribution to mimic realistic bursty behavior and is evaluated with RPS from medium load up to the cluster saturation point.
To show the effectiveness of \sysname routing, we ran it in an aggregated serving setup as well as a prefill-only serving scenario. The latter is increasingly common~\cite{prefillonly, prefill_as_service} and also serves the request scheduling for prefill instances in a Prefill-Decode(P/D) disaggregated cluster.

The cluster serves Llama3 8B model with float16 data type on vLLM v1 serving engine. In vLLM engine, chunked prefill and prefix caching were enabled, except that prefix caching was disabled on NVIDIA-V100 instances since the feature was not fully supported for it. Note that \sysname's design is not tied to any specific LLM model, engine configuration, or cluster configuration. 
For example, multi-GPU inference with tensor parallelism or pipeline parallelism can be considered as one logical instance with a group of GPUs serving each request as a whole and routing decisions made at the logical-instance level.

\paragraph{Baselines}
We compare against \heuristic (Algorithm~\ref{alg:prefix_and_load_aware_routing} in Appendix~\ref{sec:appendix_alg}), \prefixaware (Algorithm~\ref{alg:prefixaware} in Appendix~\ref{sec:appendix_alg}) with different thresholds ($\tau$), Mooncake's simulator-based routing policy (\S\ref{sec:existing_routing_policies}), and the least request load balancing policy. To evaluate adaptation, we compare \sysname against two variants: one trained only offline, and one that stops learning midway through the experiment. \sysname does not require offline training, and all the evaluation done for \sysname was learned only online without offline training.
We use TTFT as the main latency metric as discussed in \S\ref{sec:routing-policy}.


\subsection{Routing performance}
\label{sec:eval_perf}

\subsubsection{Homogeneous cluster}
\label{sec:eval_homo_cluster}
Figure~\ref{fig:ttft_mooncake_homo_aggregated} and Figure~\ref{fig:ttft_gangmukprefix_homo_aggregated} show the TTFT in Mooncake workloads and our synthetically generated different prefix sharing ratio workloads in aggregated deployment (non-P/D disaggregation). \sysname outperforms all existing heuristics regardless of the load conditions. Against \heuristic, it achieves up to $2.15\times$ lower average TTFT latency and up to $1.86\times$ lower P99 TTFT. 
Additionally, we evaluated \sysname without online learning (purple bars). It uses a routing model trained offline on the same trace for each experiment.
The gap between \sysname and \sysname without online learning shows the gap introduced by the circular dependency (\S\ref{subsec:motivation-online}). \prefixaware under any threshold setting performed worse than \sysname, revealing the fundamental limitation of heuristic rules.

Figure~\ref{fig:ttft_mooncake_homo_prefillonly} shows that \sysname also outperforms existing policies on prefill-only workloads. 
This demonstrates that \sysname is effective not only in aggregated deployment but also in prefill-only serving and in routing to P-instances in P/D-disaggregated deployments. We also evaluated TTFT performance with and without a on-the-fly quantization configuration (bitsandbytes) on vLLM in Appendix~\ref{sec:appendix_exp} Figure~\ref{fig:perf_onthefly_quant}, and \sysname achieves the best TTFT performance.

\begin{figure}[!t]
    \centering
    \includegraphics[width=0.495\textwidth]{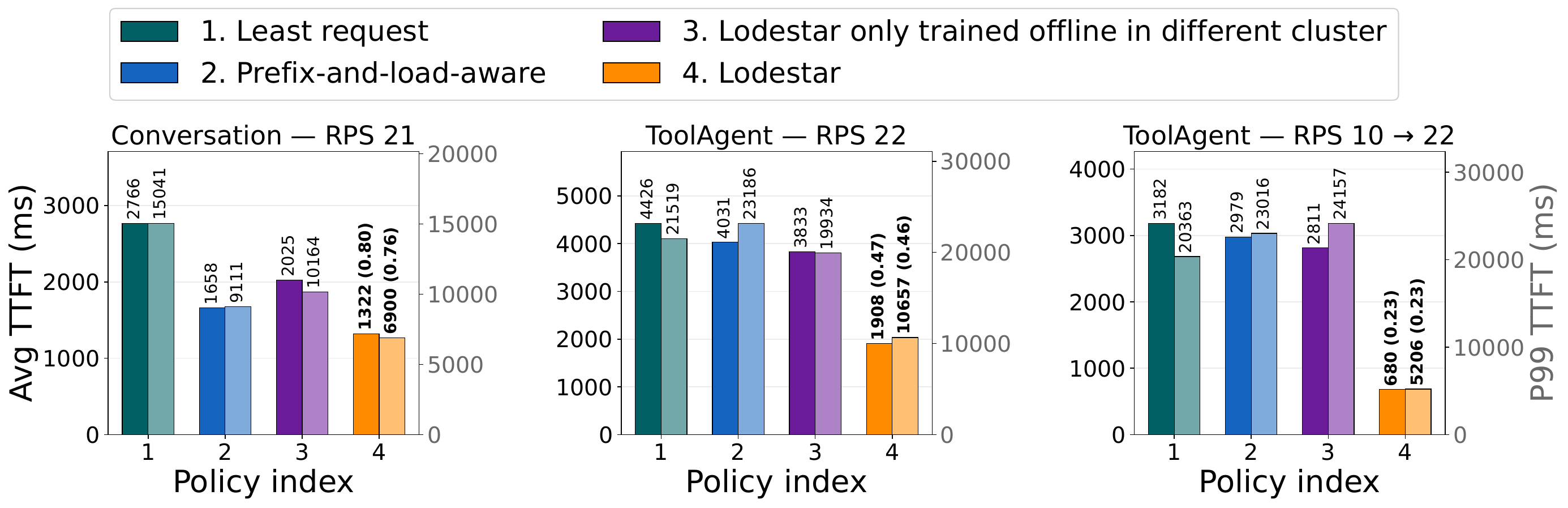}
    \caption{TTFT latency comparison in heterogeneous GPU cluster in aggregated deployment. The right-most figure shows the experiment with changing loads.}
    \vspace{-0.1in}
    \label{fig:ttft_mooncake_hetero_aggregated}
\end{figure}

\subsubsection{Heterogeneous cluster}
\label{sec:eval_hetero_cluster}

Figure~\ref{fig:ttft_mooncake_hetero_aggregated} shows the TTFT latency performance on the heterogeneous cluster that has seven NVIDIA-A30 and eight NVIDIA-V100.
\sysname consistently achieves lower average and tail latency than all baselines across workloads. Notably, we disabled vLLM's prefix caching feature on V100 because it was not fully supported on Volta architecture and instances were crashing. As shown next, the different GPU models as well as vLLM feature support lead to different latency behavior, which \sysname learns directly from data online without any hardcoded rules.

\begin{figure}[!t]
    \centering
    
    \begin{subfigure}{\columnwidth}
        \centering
        \includegraphics[width=\linewidth]{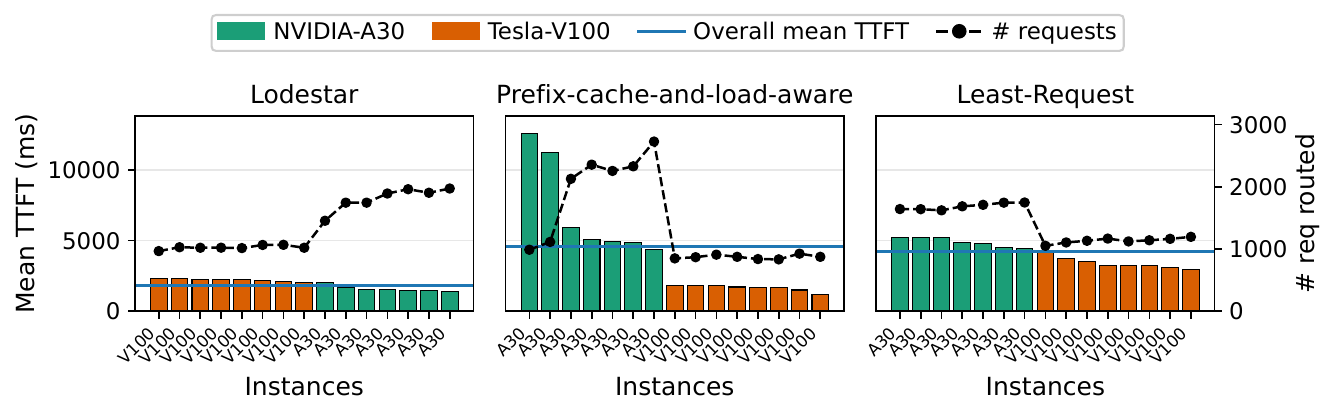}
        \caption{ToolAgent workload (RPS=22) in A30$\times$7 + V100$\times$8 cluster.}
        \label{fig:routing_hetero_toolagent}
    \end{subfigure}
    \vspace{0.5em}
    \begin{subfigure}{\columnwidth}
        \centering
        \includegraphics[width=\linewidth]{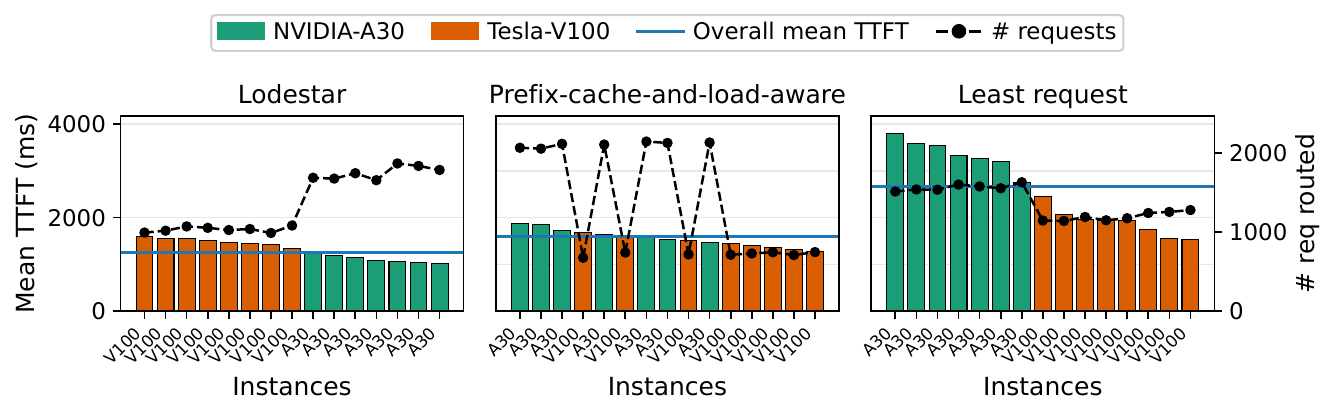}
        \caption{Conversation workload (RPS=21) in A30$\times$7 + V100$\times$8 cluster.}
        \label{fig:routing_hetero_conversation}
    \end{subfigure}
    \vspace{0.5em}
    \begin{subfigure}{0.9\columnwidth}
        \centering
        \includegraphics[width=0.78\linewidth]{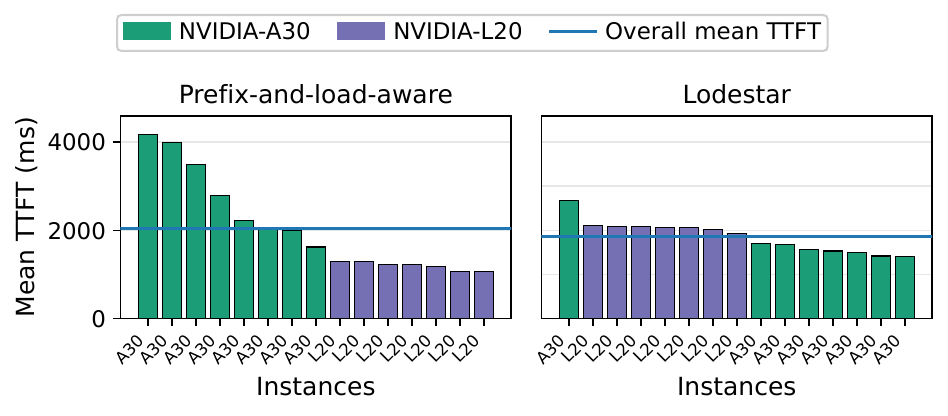}
        \caption{Conversation workload (RPS=23) in L20$\times$7 + A30$\times$8 cluster.}
        \label{fig:routing_hetero_conversation_L20x7_A30x8}
    \end{subfigure}
    \caption{Routing decision example in a heterogeneous GPU cluster for each routing policy. The bars show mean TTFT of each instance, and dots show the total number of requests routed to it. The instances are sorted from higher latency to lower from left to right on the x-axis. 
    }
    \vspace{-0.1in}
    \label{fig:routing_decision_in_hetero}
\end{figure}

Figure~\ref{fig:routing_decision_in_hetero} dissects policies' routing decisions on each GPU instance. \sysname achieves substantially lower average and tail latency.
On ToolAgent workload (Figure~\ref{fig:routing_hetero_toolagent}), \heuristic shows particularly high tail latency on two A30 instances. This is because \heuristic can create prefix hotspots. By the algorithm, it is possible that it routes requests with the highest prefix hit until the load imbalance is detected. In a worse situation, it can direct a burst of long, prefix-matching requests to the same instance, and their KV footprints accumulate quickly and saturate GPU memory, triggering cache eviction and request preemption. Recomputing the evicted KV then increases TTFT.
Least request balances request counts well; however, without prefix awareness and detailed token-level information, requests sharing the same prefix are scattered across instances. This leads to redundant prefill operations on nearly every instance and results in uniformly high latency. 
On the Conversation workload (Figure~\ref{fig:routing_hetero_conversation}), the latency result trends are similar to ToolAgent. The tail latency of \heuristic is not as bad because the larger prefix reuse distance in Conversation spreads prefix-matching requests over time, making prefix hotspots less likely to form.
Another interesting observation is that \sysname performs slightly lower latency on A30 instances (the opposite trend of the other two heuristics). 
As noted, prefix caching is enabled on A30s and disabled on V100s due to limited support of vLLM engine. \sysname learns this asymmetry from data and concentrates prefix-reusable work on A30s but only up to the point where it does not create prefix hotspots.

The learned routing policy adapts to different cluster configurations well. In a second heterogeneous cluster of seven NVIDIA-L20s and eight NVIDIA-A30s (Figure~\ref{fig:routing_hetero_conversation_L20x7_A30x8})---where both GPUs support prefix caching and the L20 offers more compute and memory than the A30 at similar memory bandwidth---\heuristic overloads the lower-end A30s, producing high tail latency on them while leaving L20s underutilized. \heuristic underestimates the A30's limited KV budget (24\,GB vs.\ 48\,GB on the L20), which can be saturated more quickly and lead to request preemption, while overestimating the benefit of the prefix cache hits it steers toward them. There is a hidden tension between memory pressure and KV cache hit benefit that static heuristics cannot navigate. \sysname finds a better policy and exhibits the opposite pattern: A30 instances see slightly lower average latency than L20s. Preemption-induced latency spikes on A30s generate a strong negative reward signal, and \sysname learns to route more conservatively to A30s in similar cluster states, lowering their latency.

Generally, \sysname correctly learns the performance of each GPU model and achieves lower latency. Note that there is no explicit modeling of GPU model performance in \sysname. It was purely learned from online data.

\begin{figure}[!t]
    \centering
    \includegraphics[width=0.49\textwidth]{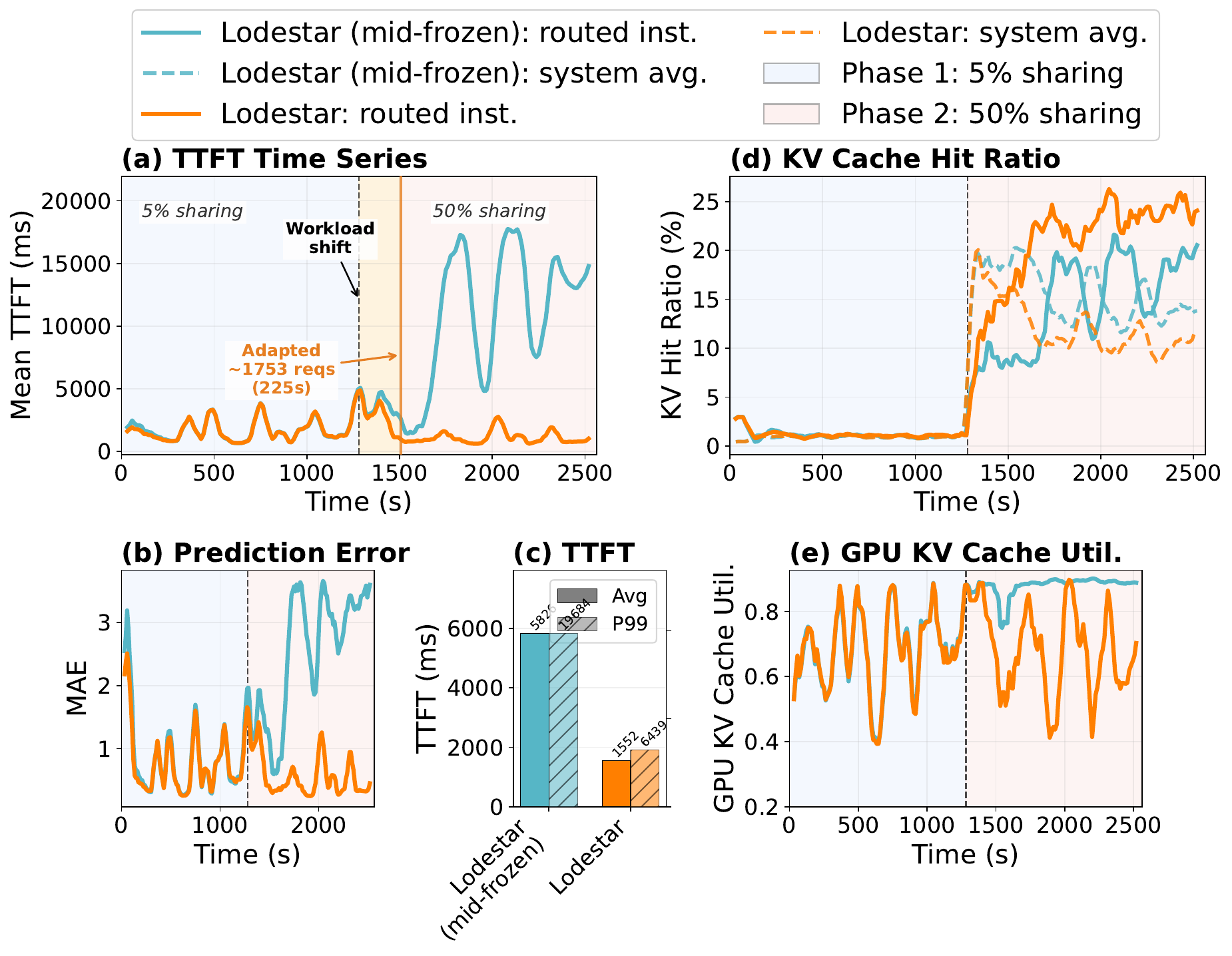}
    \caption{Online adaptation experiment with dynamically changing prefix sharing ratio distribution. The average prefix sharing ratio changes from 5\% to 50\% in the middle of the experiment. \sysname (mid-frozen) stopped learning online right before the workload changed. \sysname kept learning online continuously. In (d) and (e), the solid lines indicate the target metrics of the selected instance by routing policy. The dotted lines indicate the cluster-wise average of the metrics at the time.}
    \vspace{-0.1in}
    \label{fig:adaptation_changing_prefix_sharing_ratio}
\end{figure}

\subsection{Online adaptability}
\label{sec:eval_adaptability}
To evaluate \sysname's adaptability, we compare \emph{\sysname} and \emph{\sysname (mid-frozen)} under two dynamic workloads: a shifting prefix-sharing ratio and a shifting request rate. Figure~\ref{fig:adaptation_changing_prefix_sharing_ratio} shows the first experiment. The workload consists of 20,000 requests, and the prefix-sharing ratio rises from 5\% to 50\% at request 10,000 (Figure~\ref{fig:adaptation_changing_prefix_sharing_ratio}~(a)). Both \sysname and \sysname (mid-frozen) learn online during the first half; \sysname (mid-frozen) then freezes its model just before the shift, while \sysname continues to update.

As shown in Figure~\ref{fig:adaptation_changing_prefix_sharing_ratio}~(b) and (c), \sysname adapts to the new distribution, maintaining low prediction error and achieving lower latency, whereas \sysname (mid-frozen) fails to adapt and suffers higher prediction error and latency. The policy \sysname (mid-frozen) learns on the 1st half of the experiment is essentially focused on load balancing due to the low 5\% average prefix sharing, which is a poor fit when applied to the 50\% average prefix sharing workload. 


The third plot in Figure~\ref{fig:ttft_mooncake_hetero_aggregated} evaluates on a workload that changes request rate from 10\,RPS to 22\,RPS in the middle of the experiment. Compared to the other two static-RPS evaluations (shown in the other plots of Figure~\ref{fig:ttft_mooncake_hetero_aggregated}), \sysname achieves a larger TTFT reduction over the baselines (0.23$\times$ vs. 0.47$\times$/ 0.8$\times$), demonstrating its superior adaptability to shifting workloads. On the other hand, \sysname that only trained offline (using data collected in a different homogeneous A30 cluster) performs very poorly both because what it learns does not transfer to the current cluster environment and its inability to adapt to changing load.

Beyond adaptability, Figure~\ref{fig:adaptation_changing_prefix_sharing_ratio} (d) and (e) show a good example of the circular dependency: routing policies affect the cluster state, which affects the request latency, which in turn affects the routing policy. 
In the latter half of the 50\% sharing workload, \sysname learned that high prefix hit reduces the latency and it routes more and more requests to the high KV cache hit ratio instances. As a result of the learned policy, it specialized the distribution of KV, so a smaller number of instances have high KV cache hit ratio and other instances have less. 
At the same time, it drives the cluster-wise average KV cache hit ratio (dotted orange line) low. Also, the overall KV cache utilization in GPU memory becomes low with higher cache reuse. 
This cluster state which is the result of the learned policy reduces TTFT. And again, the model observes them and updates the model toward the direction. 
On the other hand, \sysname (mid-frozen) policy spreads out KV blocks even in the 50\%-shared workload and starts to oversubscribe KV cache space. Eventually, it leads to more KV evictions in the caches and low max KV cache hit ratio.

\begin{figure}[!t]
    \centering
    \includegraphics[width=0.43\textwidth]{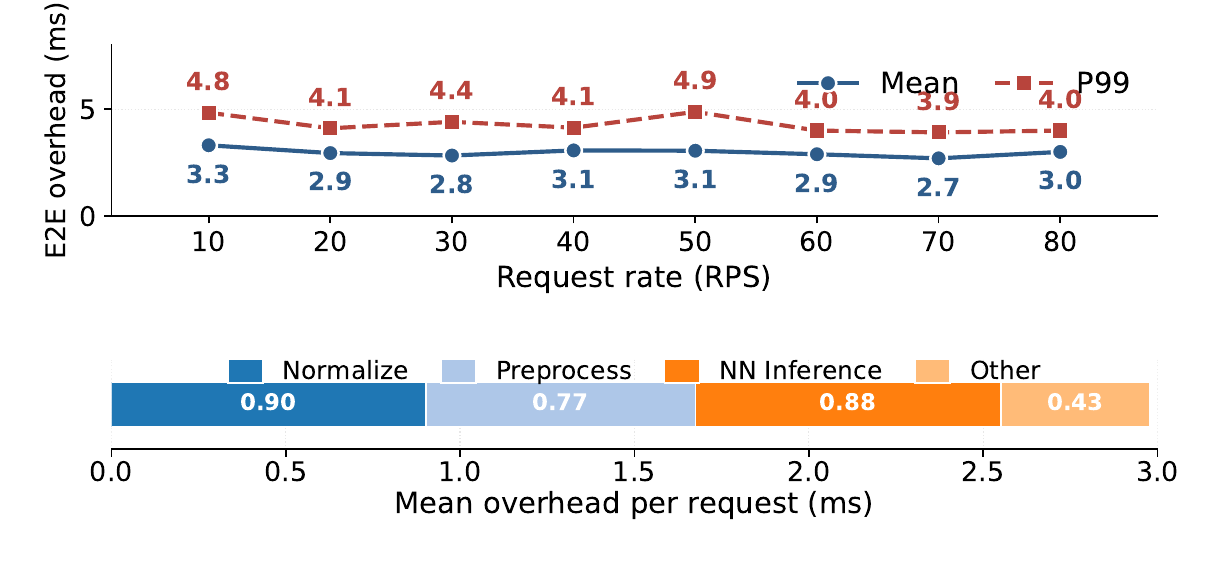}
    \vspace{-0.1in}
    \caption{\sysname overhead analysis.}
    \label{fig:overhead_analysis}
\end{figure}

\subsection{Overhead analysis}
\label{sec:eval_overhead}
Figure~\ref{fig:overhead_analysis} measures \sysname's end-to-end routing overhead on the request critical path. Sweeping request rate from 10 to 80 RPS (Figure~\ref{fig:overhead_analysis}(top)), mean overhead stays low around 3ms and tail latency around 4.5ms with no upward trend across the 8$\times$ load range.
The routing path does not degrade under load, consistent with the inference/training isolation in \S\ref{sec:design_ras} that keeps training, data flushing, and checkpointing off the hot path. 
Overall, the added overhead of \sysname pays off with a much bigger TTFT reduction. (Note that all our other experiments include the overhead in \sysname's TTFT values.)

\begin{figure}[!t]
    \centering
    \includegraphics[width=0.5\textwidth]{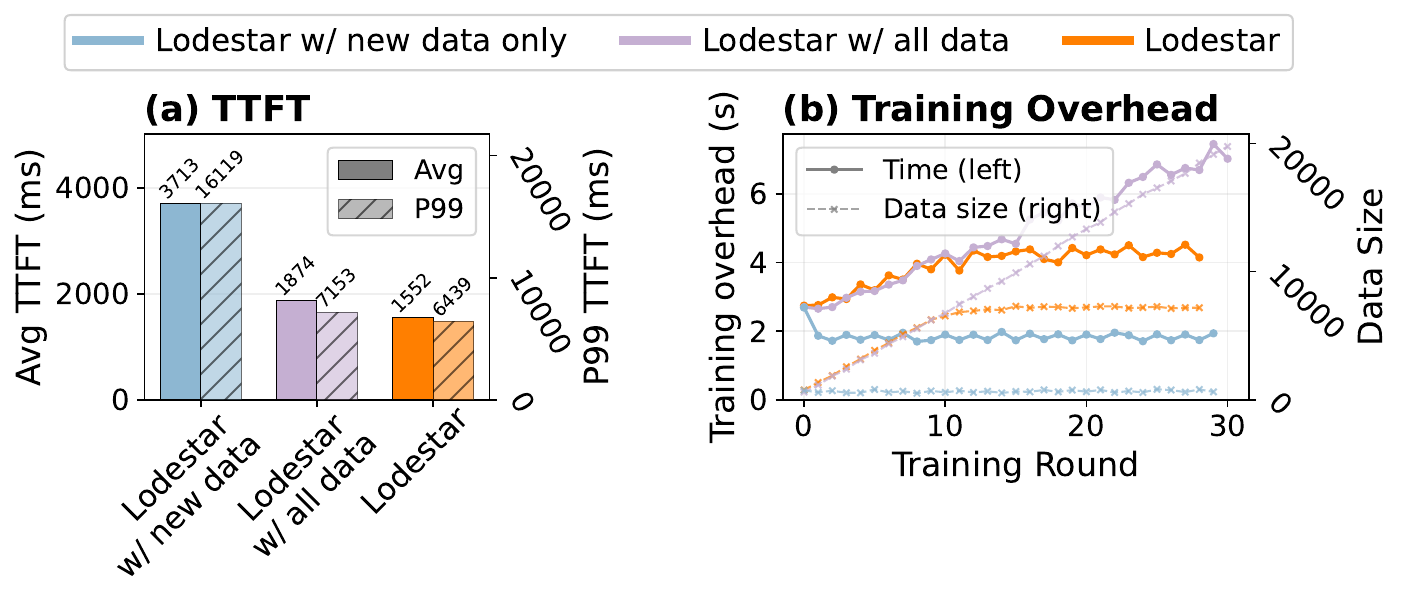}
    \caption{Training data selection algorithm ablation study (\sysnameshort: \sysname).}
    \label{fig:training_data_selection_algorithm_ablation_study}
\end{figure}

\subsection{Data selection strategy}
\label{sec:eval_data_selection_strategy}


To evaluate \sysname's two-pool data selection (FIFO buffer and replay buffer), we compare it against \emph{\sysname w/ new data only} (FIFO buffer alone) and \emph{\sysname w/ all data} (full history retained) under a workload that shifts prefix sharing ratio (5\% to 50\%) in the middle.
As Figure~\ref{fig:training_data_selection_algorithm_ablation_study}(a) shows, \sysname achieves an average/P99 TTFT that is $2.39\times$/$2.50\times$ lower than \emph{new data only} and $1.21\times$/$1.11\times$ lower than \emph{all data}. Figure~\ref{fig:training_data_selection_algorithm_ablation_study}(b) compares their training cost: \emph{w/ new data only} is the cheapest with constant overhead (FIFO buffer); \emph{all data}'s cost grows linearly as rounds accumulate; \sysname's cost initially grows linearly and plateaus once the replay buffer saturates. By retaining only informative and recent samples, \sysname's data selection is both effective and bounded in cost.

\begin{figure}[!t]
    \centering
    \includegraphics[width=0.48\textwidth]{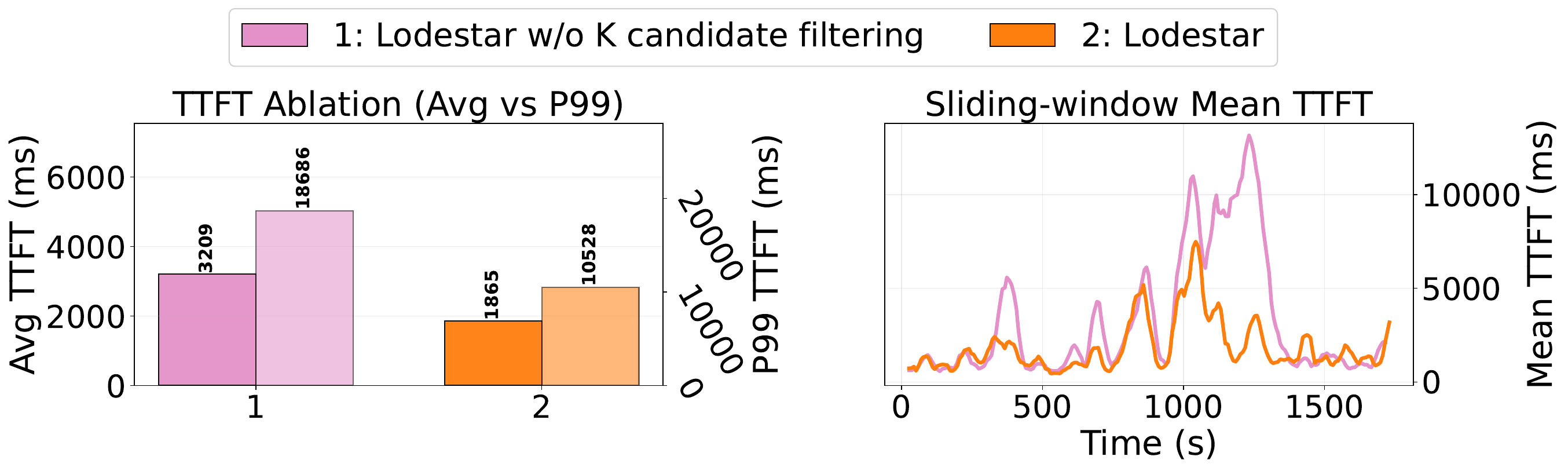}
    \caption{K candidate filtering ablation study (ToolAgent workload, RPS 12).}
    \label{fig:k_candidate_filtering_ablation_study}
    \vspace{-0.2in}
\end{figure}

\subsection{Consistent hashing based filtering}
\label{sec:eval_k_filtering}

Figure~\ref{fig:k_candidate_filtering_ablation_study} ablates K candidate filtering on the ToolAgent workload at RPS 12. Removing it raises average TTFT from 1865 to 3209\,ms ($1.72\times$) and P99 TTFT from 10528 to 18686\,ms ($1.78\times$). The sliding-window mean TTFT trace shows that both variants track closely in the early portion of the run, but the unfiltered variant has more tail excursions, entering a sustained spike between $\sim$900 and $\sim$1300\,s that peaks near 12k\,ms, while \sysname stays under $\sim$7k\,ms and recovers faster. Consistent hashing based filtering therefore contributes primarily by damping tail TTFT under load, which matches the $1.78\times$ P99 gap being larger than the $1.72\times$ mean gap.

\section{Related Work}
\label{sec:related_work}

\paragraph{Request routing \& load balancing in other applications} Prior systems pick signals matched to their workload: C3~\cite{c3} uses queue-size feedback and rate throttling for Cassandra replica selection; Prequal~\cite{prequal} probes requests-in-flight rather than CPU load at scale; Clockwork~\cite{clockwork} exploits the deterministic latency of DNN inference for centralized SLO scheduling; and SLATE~\cite{slate_hotnets} performs global traffic engineering over multi-hop microservice call graphs across geo-distributed clusters, jointly optimizing latency and egress cost. All assume stateless or near-stateless requests with uniform or tightly bounded cost, for which a small set of load- or queueing-based signals suffices to rank servers. LLM inference breaks both assumptions: per-request compute varies by orders of magnitude, and cross-request KV-cache state couples routing decisions. No fixed signal captures this mapping, motivating \sysname's learned reward predictor.

\paragraph{Intra-engine scheduling in LLM serving}
A large body of work optimizes scheduling and memory management inside a single serving engine. vLLM's PagedAttention~\cite{vllm} eliminates KV cache fragmentation, and subsequent systems extend it with dynamic or hierarchical KV storage for long contexts and multi-turn reuse~\cite{infinigen, attentionstore}. For token-level scheduling, Sarathi-Serve~\cite{sarathi-serve} interleaves chunked prefill with decode to mitigate head-of-line blocking, NanoFlow~\cite{nanoflow} pipelines intra-device compute, memory, and network operations, and FastServe~\cite{fastserve}, learning-to-rank SJF~\cite{learning_to_rank}, and JITServe~\cite{jitserve} provide preemptive, length- and SLO-aware ordering under imprecise request-length information. 

\paragraph{Other LLM inference optimizations} 
Complementary efforts optimize other dimensions of LLM serving: speculative decoding~\cite{speculative_decoding, medusa, eagle}, quantization and sparsity for memory efficiency~\cite{flashattention}, and memory offloading or GPU sharing~\cite{xing2025towards, yu2025prism}. Emerging disaggregated LLM architectures~\cite{distserve, dynaserve, deserve, megascale-infer} and KV-cache compression or memory-tiering approaches aim to scale serving efficiency further. \sysname will automatically adapt to whatever optimizations are applied to the serving system.

\paragraph{Semantic router}
Semantic routers~\cite{frugalgpt, llmrank, learningroutellms, moma, vllmsemanticrouter} pick \emph{which} model or agent to serve a query, trading cost against quality based on query characteristics. \sysname is orthogonal: it operates after model selection, routing to specific instances of the same model based on runtime cluster state.




\section{Conclusion}
We presented \sysname, a novel learning-based request routing system for LLM inference applications. \sysname learns request routing policy adaptively directly from data without requiring explicit white-box modeling or simulation. It is able to continuously learn the latency behavior with different request information, KV state, instance state in dynamic workloads and infrastructure conditions. \sysname outperforms existing LLM inference request routing systems by achieving significantly lower average latency and P99 latency.



\bibliographystyle{plain}
\bibliography{references}



\clearpage
\appendix

\section{Existing Routing Algorithms}
\label{sec:appendix_alg}

\begin{algorithm}[th]
\small
\begin{algorithmic}[1]
\Require Request $r$; ready instances $\mathcal{P}$; prefix match ratios $\{m_i\}_{i \in \mathcal{P}}$; request counts $\{c_i\}_{i \in \mathcal{P}}$; imbalance threshold $\theta_{\text{imbal}}$; overload factor $k$
\Ensure Selected instance $i^*$
\State $c_{\min}, c_{\max} \gets \min_i c_i,\; \max_i c_i$
\If{$c_{\max} - c_{\min} > \theta_{\text{imbal}}$} \Comment{Load imbalanced}
    \State \Return $\arg\min_i c_i$ \Comment{Least-loaded instance}
\EndIf
\State $\mu, \sigma \gets \textsc{Mean}(\{c_i\}),\; \textsc{StdDev}(\{c_i\})$
\State $\mathcal{S} \gets \text{sort } \mathcal{P} \text{ by } (m_i \text{ desc},\; c_i \text{ asc})$
\For{$i \in \mathcal{S}$}
    \If{$c_i \leq \mu + k \cdot \sigma$} \Comment{best prefix match not overloaded}
        \State \Return $i$
    \EndIf
\EndFor
\State \Return $\arg\min_i c_i$ \Comment{Fallback: least-loaded}
\end{algorithmic}
\caption{\heuristic routing policy}
\label{alg:prefix_and_load_aware_routing}
\end{algorithm}

\begin{algorithm}[th]
\small
\begin{algorithmic}[1]
\Require Request $r$; ready instances $\mathcal{P}$; prefix match ratios $\{m_i\}_{i \in \mathcal{P}}$; request counts $\{c_i\}_{i \in \mathcal{P}}$; prefix hit threshold $\tau$
\Ensure Selected instance $i^*$
\State $i^* \gets \arg\max_{i \in \mathcal{P}} m_i$
\If{$m_{i^*} > \tau$}
    \State \Return $i^*$ \Comment{highest prefix match, above threshold}
\EndIf
\State \Return $\arg\min_{i \in \mathcal{P}} c_i$ \Comment{Fallback: least-loaded}
\end{algorithmic}
\caption{\prefixaware routing policy with threshold $\tau$}
\label{alg:prefixaware}
\end{algorithm}

\section{\sysname Algorithms}
\label{sec:appendix_qs} 

\begin{algorithm}[!h]
\caption{Routing on \gatewayfull}
\label{alg:routing-with-fallback}
\small
\begin{algorithmic}[1]
\Require Request features $\mathbf{r}$; instance states $\{\mathbf{x}_i\}_{i=1}^{N}$; KV hit ratios $\{\kappa_i\}_{i=1}^{N}$; heuristic policy $\mathcal{H}$ (\heuristic)
\Ensure Request forwarded to selected instance $i^*$
\State $i_{\text{heur}} \gets \mathcal{H}(\mathbf{r},\{\mathbf{x}_i\},\{\kappa_i\})$
  \Comment{pre-compute so fallback adds no latency}
\State $(i^*, \text{status}) \gets \textsc{Infer}(\mathbf{r},\{\mathbf{x}_i\},\{\kappa_i\})$ on \ras with timeout
\If{timeout \textbf{or} error}
  \State $i^* \gets i_{\text{heur}}$ \Comment{gateway-side slowdown/failure fallback}
\ElsIf{status $\neq$ ok}
  \State $i^* \gets i_{\text{heur}}$ \Comment{\ras-signaled fallback (cold-start or OOD)}
\EndIf
\State Forward request to instance $i^*$
\end{algorithmic}
\end{algorithm}

\begin{algorithm}[!t]
\caption{Routing logic in \ras}
\label{alg:ras-infer}
\small
\begin{algorithmic}[1]
\Require Request features $\mathbf{r}$; instance states $\{\mathbf{x}_i\}_{i=1}^{N}$; KV hit ratios $\{\kappa_i\}_{i=1}^{N}$; reward model $f_\theta$; feature bounds $[\mathbf{l},\mathbf{u}]$; collected sample count $n$; cold-start threshold $n_{\min}$; exploration rate $\varepsilon$; K-filter thresholds $(\tau_{\text{sat}}, \tau_{\text{ben}})$; tiebreak margin $\delta$
\Ensure $(i^*, \text{status})$
\If{$n < n_{\min}$} \Return $(\bot, \text{cold-start})$ \Comment{not enough samples to train}
\EndIf
\If{$\exists\, j:\ \text{feat}_j(\mathbf{r},\{\mathbf{x}_i\}) \notin [l_j, u_j]$}
  \State \Return $(\bot, \text{OOD})$ \Comment{per-feature extrapolation guard}
\EndIf
\If{$\text{rand}() < \varepsilon$}
  \State \Return $(\textsc{Uniform}(\{1, \ldots, N\}), \text{ok})$ \Comment{$\varepsilon$-greedy exploration; skip MLP}
\EndIf
\State $\tilde{\mathbf{x}}_i, \tilde{\kappa}_i, \tilde{\mathbf{r}} \gets \textsc{Normalize}(\mathbf{x}_i, \kappa_i, \mathbf{r})$
  \Comment{per-feature z-score}
\State $\hat{y}_i \gets f_\theta(\tilde{\mathbf{x}}_i, \tilde{\kappa}_i, \tilde{\mathbf{r}}), \ \forall i \in \{1, \ldots, N\}$
  \Comment{single batched call; shared $f_\theta$ across all $N$ instances}
\State $i^* \gets \arg\max_i \hat{y}_i$
\If{$\overline{\text{gpu\_kv}} > \tau_{\text{sat}}$ \textbf{and} $\max_i \kappa_i \cdot |\mathbf{r}| > \tau_{\text{ben}}$}
  \State $\mathcal{C} \gets \textsc{ConsistentHash}_K(\text{prefix\_group}(\mathbf{r}))$
  \If{$i^* \notin \mathcal{C}$}
    \State $i^* \gets \arg\max_{i \in \mathcal{C}} \hat{y}_i$ \Comment{K-filter, \S\ref{sec:routing-policy}}
  \EndIf
\EndIf
\If{$|\{i : \hat{y}_i \geq (1-\delta)\, \max_j \hat{y}_j\}| > 1$}
  \State $i^* \gets \textsc{Uniform}$ over near-best set \Comment{reward tiebreak}
\EndIf
\State \Return $(i^*, \text{ok})$
\end{algorithmic}
\end{algorithm}

\clearpage

\section{Additional Experimental Details and Results}
\label{sec:appendix_exp}

\begin{figure}[ht]
    \centering
    \includegraphics[width=0.48\textwidth]{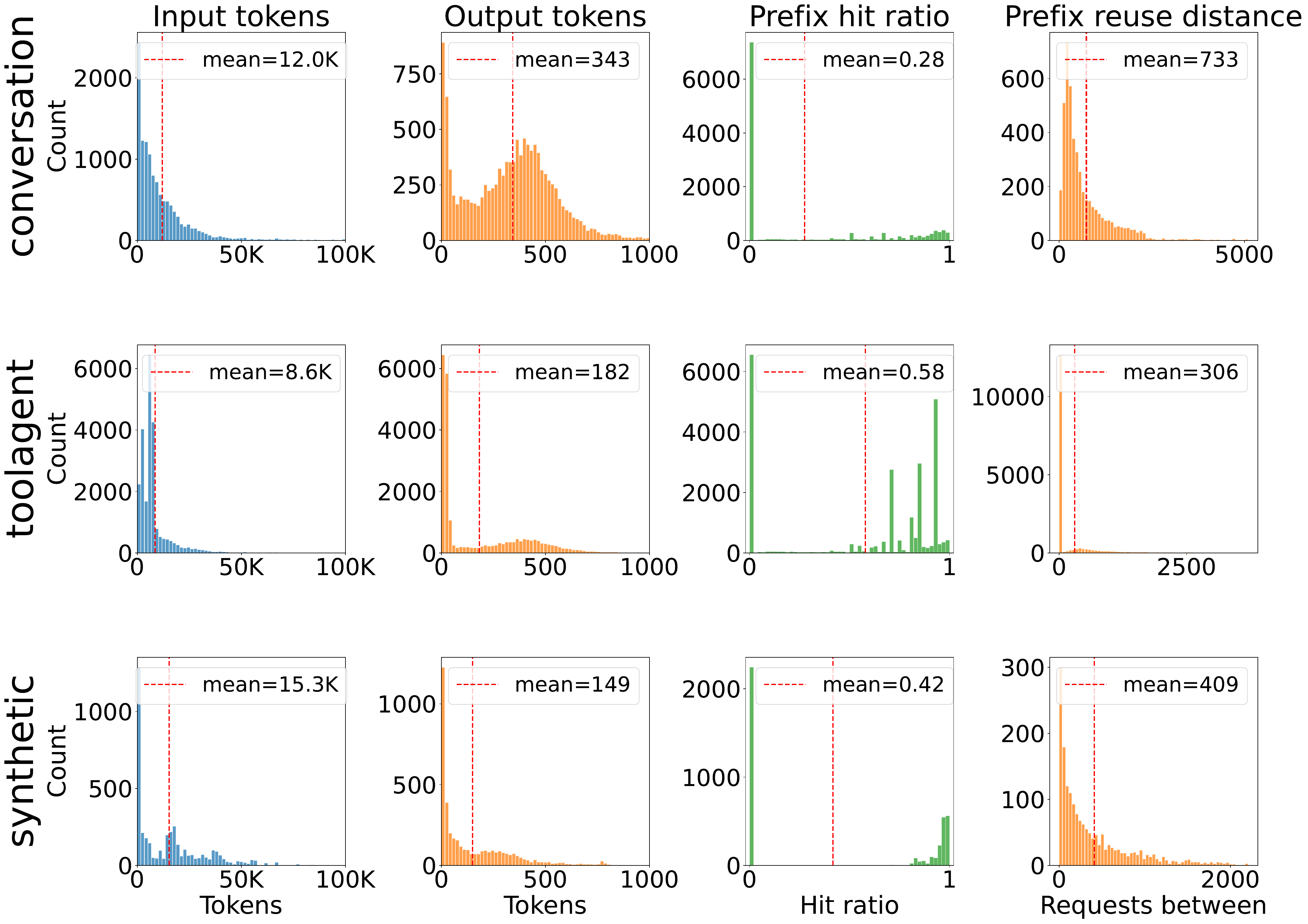}
    \caption{Mooncake workload distribution (Conversation, ToolAgent, Synthetic).}
    \label{fig:mooncake_workloads}
\end{figure}

\begin{figure}[ht]
    \centering
    \includegraphics[width=0.48\textwidth]{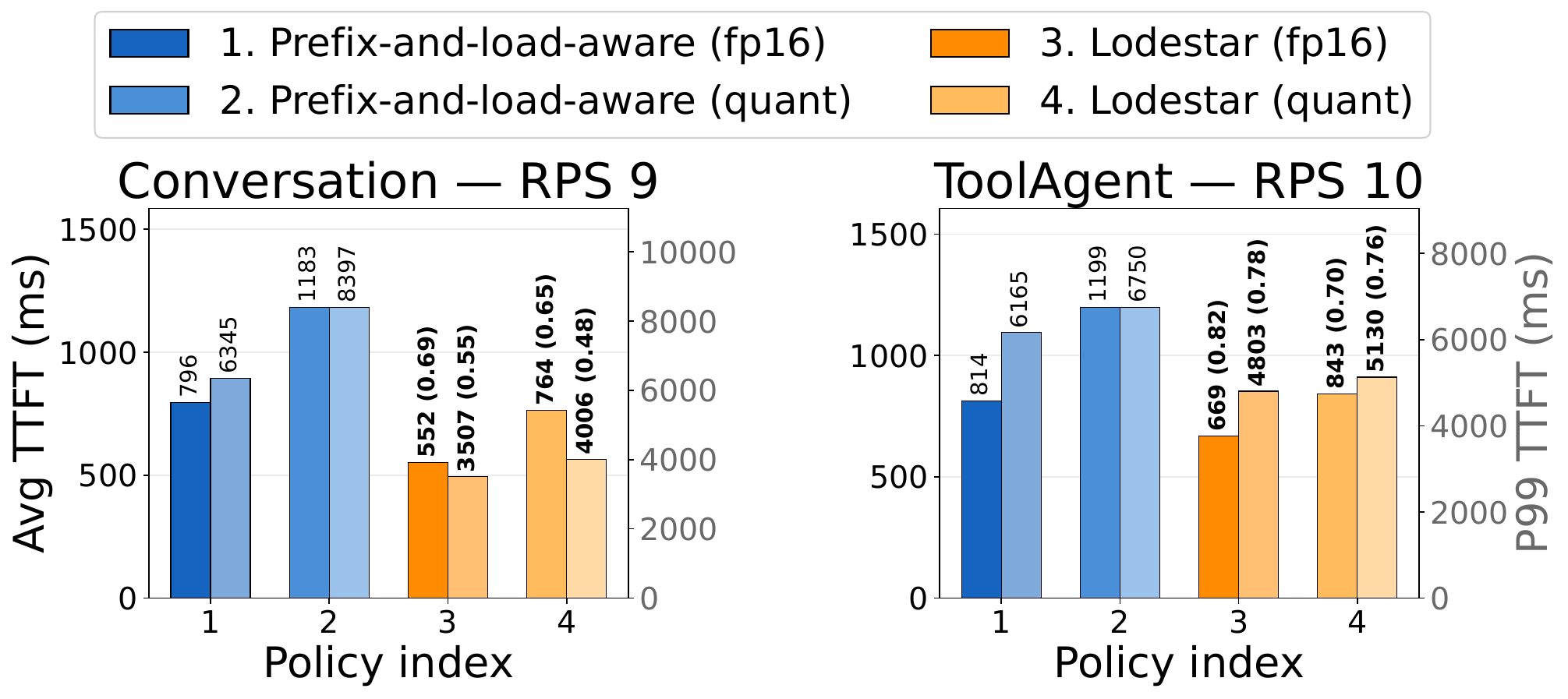}
    \caption{TTFT performance with and without bitsandbytes quantization configuration on vLLM. bitsandbytes is a popular on-the-fly quantization method that compress FP16 KV into INT4/INT8 when storing them. When it is used, vLLM engine decompresses the KVs back to the original precision. It enables memory-efficient LLM inference at the cost of computation overhead during decompress.}
    \label{fig:perf_onthefly_quant}
\end{figure}


\end{document}